\documentclass[11pt,psfig]{article}
\usepackage{graphicx}
\usepackage{float}
\usepackage{subcaption}
\usepackage{amsmath,amssymb,mathtools,bbm, amsopn}
\usepackage{tipa}
\usepackage{esint}
\usepackage{bbold}
\usepackage{color}
\usepackage{cite}
\usepackage{tikz}
\usetikzlibrary{shapes}
\usepackage{longtable}
\usepackage{pgfplots}
\usepackage{subeqnarray}
\usepackage{dsfont}
\usepackage{changepage}
\usepackage{algorithmic}
\usepackage{algorithm}
\usepackage{bigints}
\usepackage{caption}
\usepackage{amsthm}
\usepackage{enumerate}
\usepackage{amsfonts}

\tikzset{cross/.style={cross out, draw=black, minimum size=2*(#1-\pgflinewidth), inner sep=0pt, outer sep=0pt},
cross/.default={1pt}}

\usepackage[symbol]{footmisc} %footnote

\begin{document}
%\maketitle 
\begin{center}

\begin{Large} 
\noindent Comparison on two ways of phonon spectral property reconstruction from thermal spectroscopy experiment \\
\end{Large}

\vspace{0.7cm}
\begin{small}
\noindent Mojtaba Forghani$^1$\footnote[1]{email: mojtaba@stanford.edu} and Nicolas G. Hadjiconstantinou$^{2}$ \\
\end{small}

\vspace{0.3cm}
\begin{footnotesize}
$^1$ Department of Mechanical Engineering, Stanford University, CA \\
$^2$ Department of Mechanical Engineering, Massachusetts Institute of Technology
\end{footnotesize}

\end{center}

\begin{abstract}
We examine the reliability of the cumulative thermal conductivity as a function of the free path, $F(\Lambda)$, in the context of reconstruction of phonon transport properties from thermal spectroscopy experiments. We specifically show that a given $F(\Lambda)$ {\it  does not} correspond to a unique relaxation-times function, in the sense that more than one distribution of relaxation-times can result in the same $F(\Lambda)$. Since different relaxation-time distributions will, in general, lead to different thermal responses, $F(\Lambda)$ does not uniquely predict the material thermal response in all transport regimes. This implies that in the context of thermal transport at the nanoscale within the Boltzmann relaxation-time approximation framework, the ``fundamental" property that should be reconstructed from the thermal spectroscopy experiments is the frequency-dependent relaxation-times function (provided group velocities are known), since it explicitly appears in the governing equation as the input material property. Extensive global optimization studies show that a previously proposed formulation for reconstruction based on the frequency-dependent relaxation-times function [Physical Review B 94, 155439 (2016)] provides numerically unique solutions. 
\end{abstract}

\section{Introduction}
\label{intro}
The study of phonon dynamics in the context of nanoscale solid-state heat transport has received considerable attention \cite{ThermalSpectroscopy2011, Mojtaba2016, Mojtaba2018, MojtabaAPL, Cahill2001, Johnson, Minnich2012, Broido2010, Esfarjani2011, Pop, Pop_micro, NanoTransport2, Tian, Zebarjadi}. One area of significant interest is the study of phonon relaxation-time and free path distributions \cite{McGaughey2010, ThermalSpectroscopy2011, Maznev2011, McGaughey, Dames2015, Minnich2012, Lingping_nature, Pop_micro}. This information is required for modeling heat transport at the kinetic level, which becomes necessary due to the failure of Fourier-based analyses at such small scales. The applications of this study include improved heat management in nanoelectronic circuits and devices \cite{Yu_nature, Wingert, Pop, Pop_micro, Cahill2001, Cahill1997}, microelectromechanical sensors \cite{NanoTransport2} and nano-structured materials for improved thermoelectric conversion efficiency \cite{Biswas, Hochbaum, Boukai, Tian, Zebarjadi, Kraemer}. 

\begin{figure}[htbp]
%\vspace{-1cm}
\hspace{-0.5cm}
\centering
\begin{subfigure}{.49\textwidth}
\centering
\includegraphics[width=1.0\linewidth]{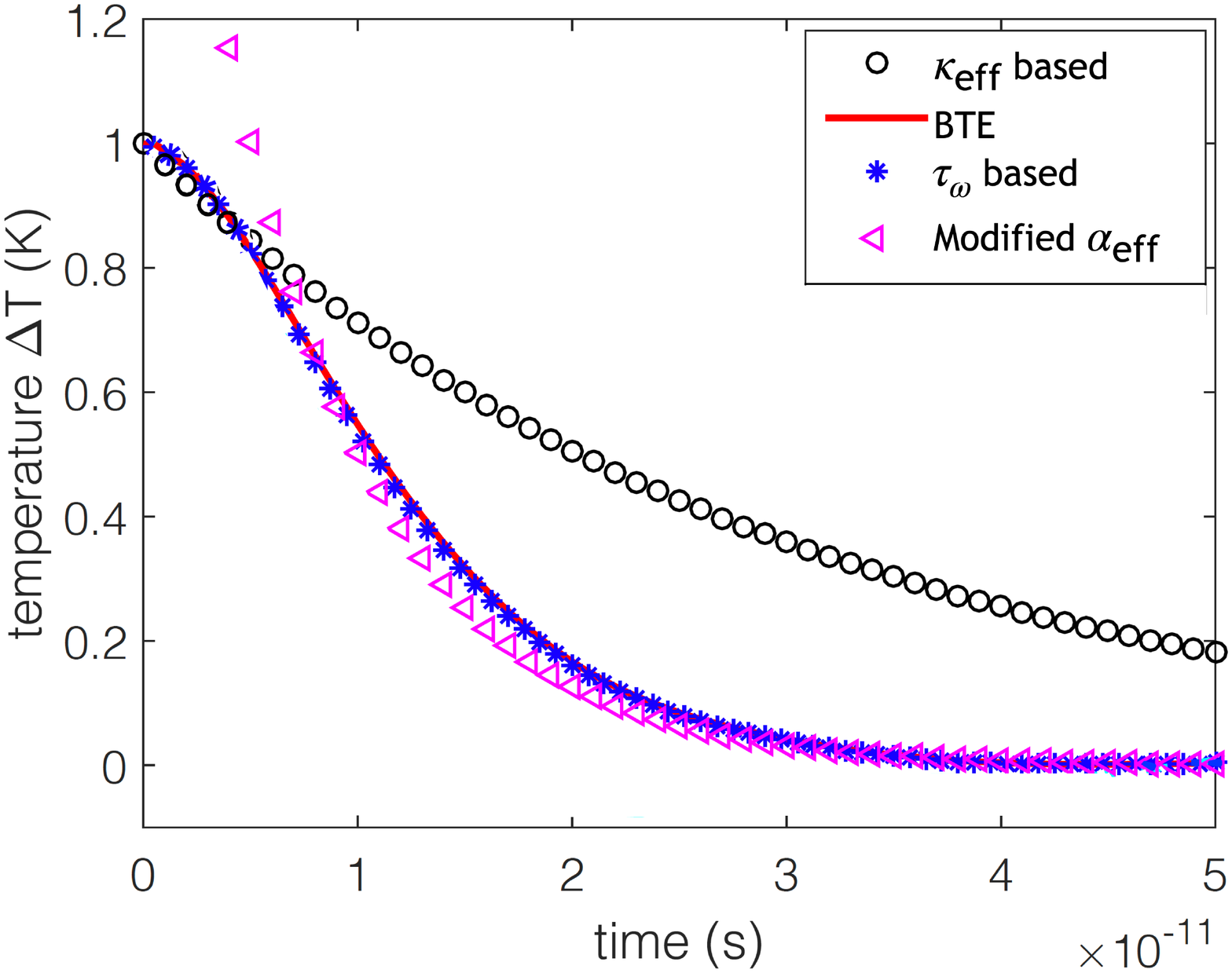}
\caption{100 nm}
\label{ab_initiomodel}
\end{subfigure}
\begin{subfigure}{.49\textwidth}
\centering
\includegraphics[width=1.0\linewidth]{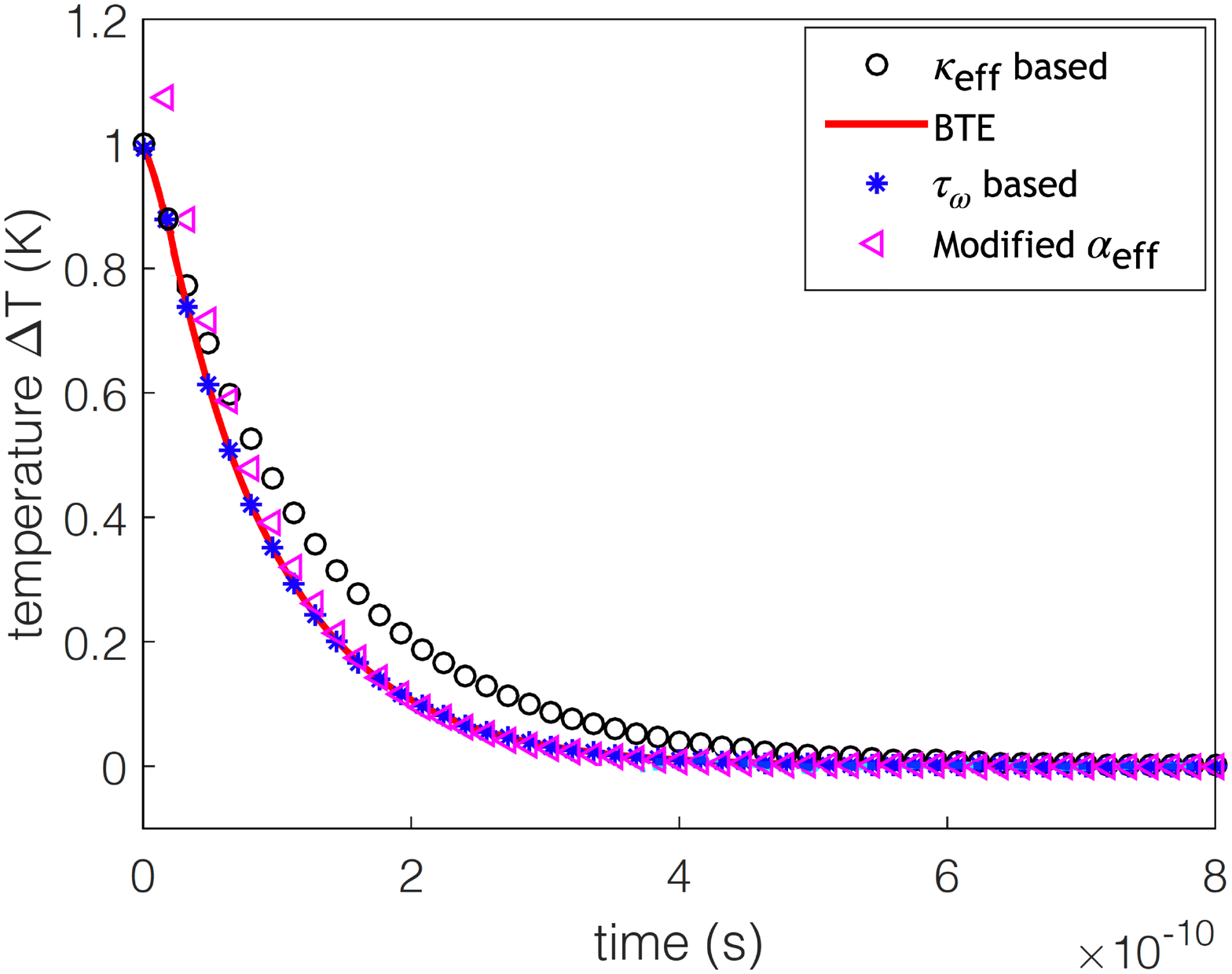}
\caption{400 nm}
\label{Hollandmodel}
\end{subfigure}
\caption{Thermal response in 1D-TTG experiment for 100 nm (a) and 400 nm (b) grating sizes. Comparison between the Boltzmann transport equation solution (denoted ``BTE''), Boltzmann transport equation solution with $\tau_\omega$ reconstructed from the BTE solution using the optimization method proposed in \cite{Mojtaba2016, Mojtaba2018} and further analyzed here (denoted by ``$\tau_\omega$ based''), Fourier heat conduction solution with effective thermal conductivity  calculated using the well-known suppression function for a thermal grating~\cite{Hua2014, Johnson} (denoted by ``$\kappa_\text{eff}$ based'') and exponential response obtained from solution of Boltzmann transport equation \cite{Mojtaba2016} (denoted by ``Modified $\alpha_\text{eff}$'').}
\label{k_eff_issue}
\end{figure}

Purely theoretical approaches such as density functional theory (DFT) have been widely used to predict phonon properties \cite{Broido2010, Esfarjani2011}. However, such approaches have yet to rise to a stature level where they can replace experimental results, in part due to their failure to consistently reproduce experimental observations \cite{Esfarjani2011}. Another popular (alternative) class of approaches relies on extracting material constitutive information from thermal spectroscopy experiments \cite{ ThermalSpectroscopy2011, Mojtaba2016, Mojtaba2018, MojtabaAPL, Johnson, Minnich2012, Zebarjadi}. However, the analysis of thermal spectroscopy data remains a challenging task. One typical approach consists of extracting the cumulative distribution function (CDF) of thermal conductivity as a function of the free path, $F(\Lambda)$ (see section~\ref{Governing_equ} for more detail), from the experimentally measured temperature relaxation profiles, by invoking the concept of ``effective thermal conductivity" and proceeding to match the experimentally measured response to solutions of the heat conduction equation with the thermal conductivity (or thermal diffusivity) treated as an adjustable, ``effective", property \cite{Lingping_nature, ThermalSpectroscopy2011}. Unfortunately, as it was discussed extensively before~\cite{Mojtaba2016}, this procedure implicitly assumes that heat transport is Fourier-like, which is only justified under fairly restrictive conditions (late times and large scales) that are not always satisfied under experimental conditions. This is highlighted in figure \ref{k_eff_issue}, adapted from \cite{Mojtaba2016} (figure 13 of \cite{Mojtaba2016}). The figure shows calculations for the one-dimensional transient thermal grating (1D-TTG) geometry~\cite{Mojtaba2016} in Si material for two grating sizes. In both cases, the reconstruction based on effective thermal conductivity approach (denoted by ``$k_\text{eff}$-based'') predicts an inaccurate thermal response compared to the one predicted by the methodology proposed in \cite{Mojtaba2016} and further analyzed here (denoted by ``$\tau_\omega$-based''). The true thermal response is denoted by ``BTE''(Boltzmann transport equation)---see section \ref{Governing_equ} for more detail. It is clear that the error in the $k_\text{eff}$-based reconstruction increases as the length scale decreases.

\begin{figure}[htbp]
%\vspace{-1cm}
\hspace{-0.5cm}
\centering
\begin{subfigure}{.49\textwidth}
\centering
\includegraphics[width=1.0\linewidth]{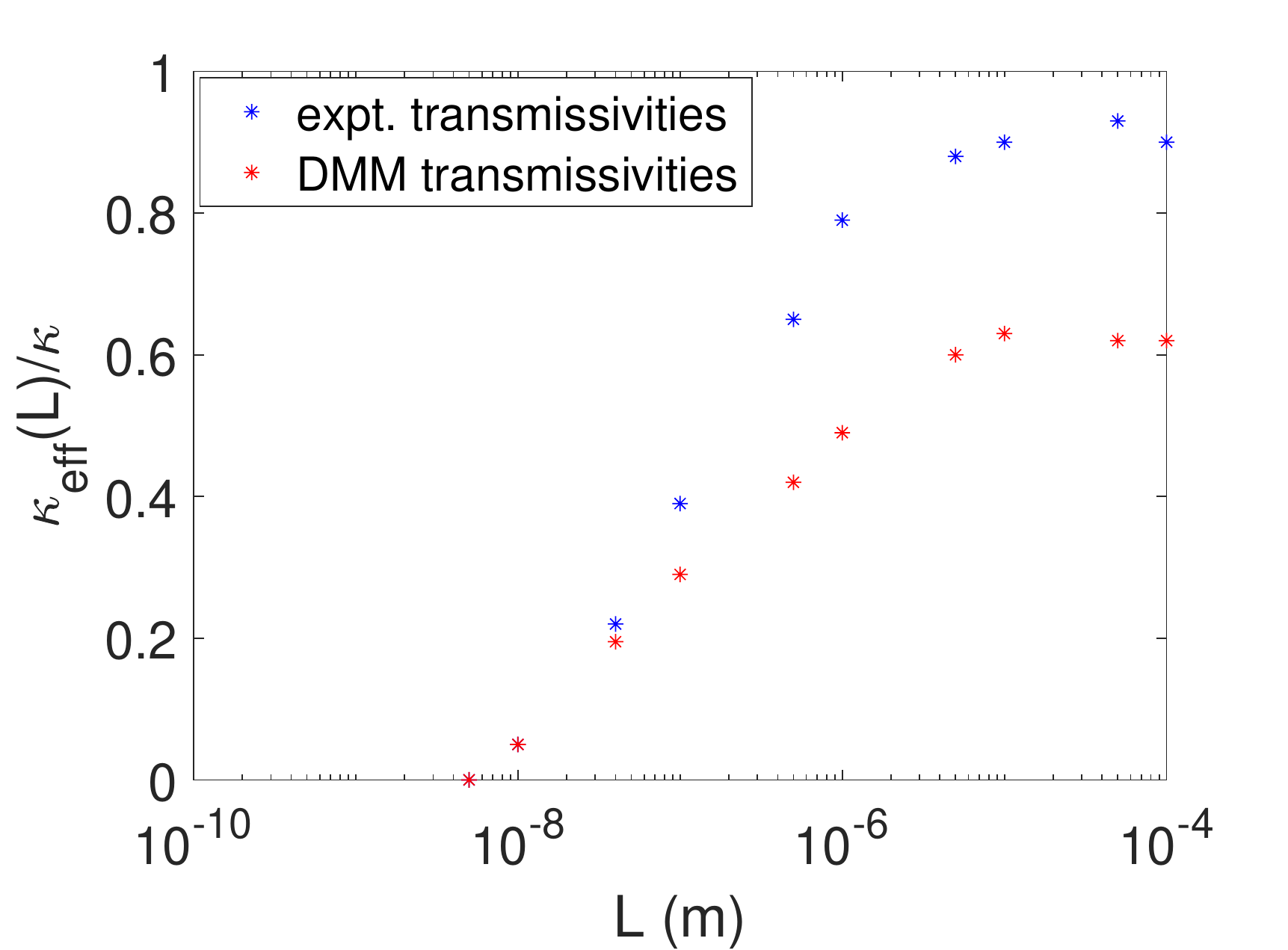}
\caption{}
\label{Hollandmodel}
\end{subfigure}
\begin{subfigure}{.49\textwidth}
\centering
\includegraphics[width=1.0\linewidth]{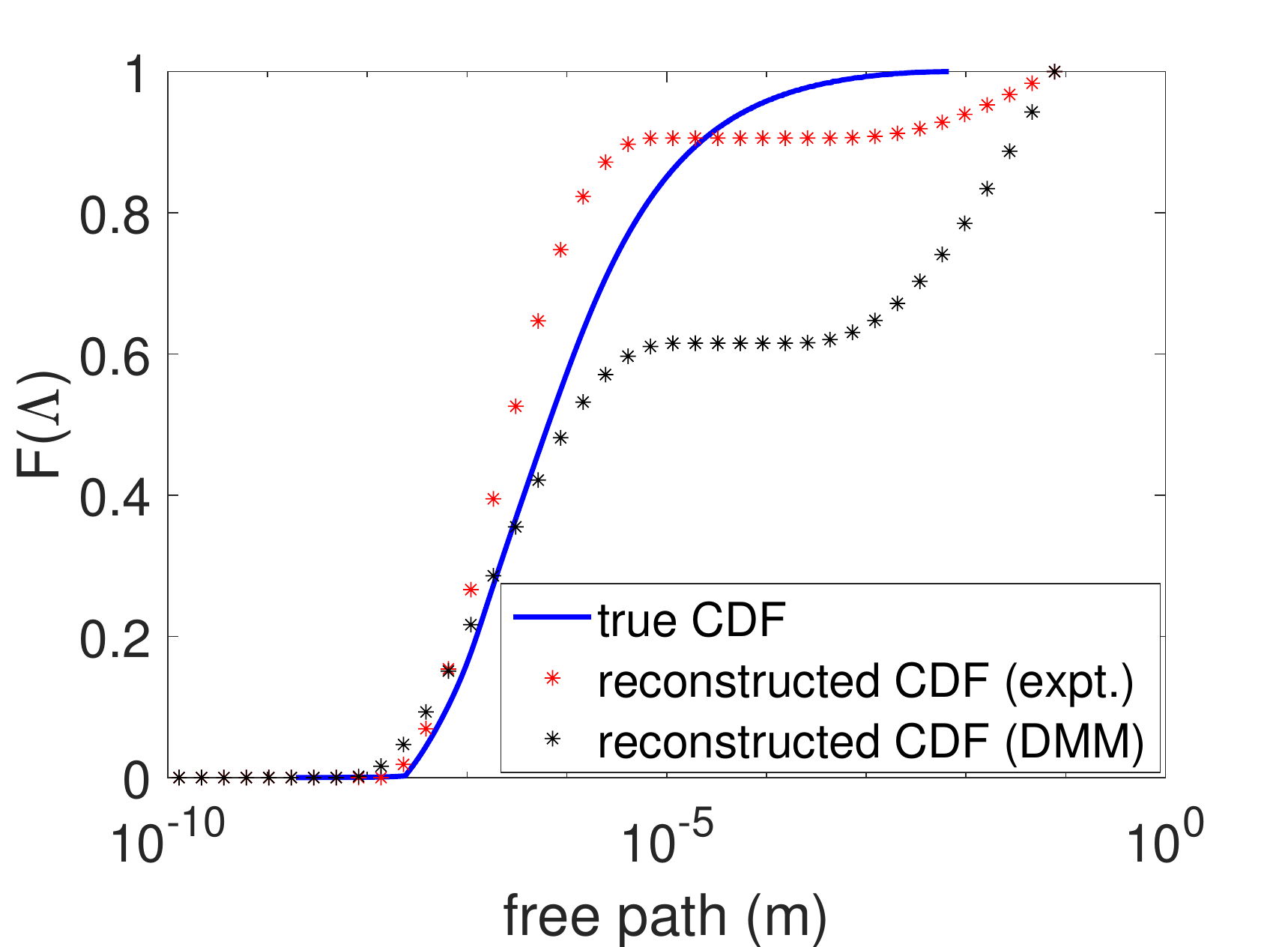}
\caption{}
\label{T_100n_unique}
\end{subfigure}
\caption{Reconstruction results using the effective thermal conductivity approach for the 2D-dots geometry, based on synthetic data generated by solution of the BTE \cite{Mojtaba2018}. The effective thermal conductivities do not match if different interface models (DMM denotes diffuse mismatch model; expt. denotes experimentally determined transmissivities~\cite{expt_trans}) are used when generating the synthetic data (a).  As a result, the corresponding $F(\Lambda)$ do not match and are each different from the true $F(\Lambda)$ (b).}
\label{k_eff_issue_2}
\end{figure}

Figure \ref{k_eff_issue_2} shows examples of reconstructing Si properties in a 2D-dots geometry~\cite{Lingping_nature} from synthetic (numerical) data, adapted from \cite{Mojtaba2018} (figure 10 of \cite{Mojtaba2018}). The two $\kappa_\text{eff}(L)/\kappa$ plots (figure \ref{k_eff_issue_2}a) correspond to the reconstructed cumulative effective heat conductivity functions, as a function of the system length scale, based on two different Al-Si interface models (transmissivity functions) that have been used in the numerical experiment (Monte-Carlo solution of the BTE); neither is able to recover the material thermal conductivity (reach the value of 1). This is due to experiment not satisfying the late time and large scale criteria needed for the $\kappa_\text{eff}$-based approach to be valid (see \cite{Mojtaba2018} for more detail). Consequently, the two reconstructed $F(\Lambda)$ shown in  figure \ref{k_eff_issue_2}b, corresponding to reconstructions based on these effective heat conductivities, have not been able to recover the true CDF.

As noted above, in response to the failure of $\kappa_\text{eff}$-based methods to provide accurate reconstructions at all scales, in \cite{Mojtaba2016, Mojtaba2018, MojtabaAPL} we developed and extensively validated an optimization-based methodology for directly reconstructing phonon relaxation-times from thermal spectroscopy data. In this work, we attempt to address some more fundamental questions, such as the uniqueness of reconstructed quantities (e.g., $F(\Lambda)$), as well as their ability to reproduce the experimental results. In particular, we show that reconstructions that use $F(\Lambda)$ (as a means to defining ``effective thermal conductivities") fail to guarantee a unique thermal response. This can be understood by noting that it is the relaxation-time and {\it not} $F(\Lambda)$ which appears as a primary material-parameter input to the BTE. Specifically, we show that the same $F(\Lambda)$ can be arrived at from multiple relaxation-time distributions, which, in general, will correspond to multiple thermal responses. This observation seriously questions the reliability of approaches which use $F(\Lambda)$ in the reconstruction.

We also study the reliability of the solutions returned by the optimization algorithm proposed in \cite{Mojtaba2016} and in particular the solution uniqueness and sensitivity to noisy measurements. Our results show that the optimized solution exhibits  properties associated with a unique minimum; furthermore, the solution is sufficiently robust to noisy measurements.

The remainder of the paper is organized as follows. In section \ref{Governing_equ}, we briefly overview the governing equations, the linearized BTE and related conservation laws. In section \ref{free_path_material}, we show theoretically and numerically that more than one relaxation-times distribution, and consequently material thermal responses, can be related to the same $F(\Lambda)$, and hence, the latter may not be used to study thermal behaviors in all regimes. In section \ref{Unique_analysis}, we investigate the uniqueness of the solution associated with our previously proposed algorithm which instead reconstructs the relaxation-times function; we also investigate the sensitivity of this approach to the uncertainty in the temperature measurements from a Bayesian perspective. Finally, in section \ref{Conclude} we provide a summary of our work and suggestions for future improvements.

\section{Boltzmann transport equation}
\label{Governing_equ}
Given the small temperature differences usually associated with thermal spectroscopy experiments, here we consider the linearized BTE 
\begin{equation} \label{BTE}
\frac{ \partial e^d } { \partial t } + \textbf{v}_{\omega} \cdot \nabla_{\textbf{x}} e^d = -\frac{ e^d- (de^\text{eq}/dT)_{T_\text{eq}} \Delta \widetilde{T} } { \tau_{\omega} } ,
\end{equation}
where $e^d= e^d (t, \mathbf{x}, \omega, \mathbf{\Omega})= e-e^\text{eq}_{T_\text{eq}}= \hbar \omega (f- f^\text{eq}_{T_\text{eq}}) $ is the deviational energy distribution, $\omega$ is the phonon frequency, $\mathbf{\Omega}$ is the phonon traveling direction, $f= f (t, \mathbf{x}, \omega, \mathbf{\Omega})$ is the occupation number of phonon modes, $\textbf{v}_{\omega}= \textbf{v} (\omega)$ is the phonon group velocity, $\tau_{\omega}= \tau (\omega)$ is the frequency-dependent relaxation-time, also referred to here as the ``relaxation-times function", and $\hbar$ is the reduced Planck constant. Here and in what follows, we use $\omega$ to denote the dependence on both frequency and polarization.

The above equation is linearized about the equilibrium temperature $T_\text{eq}$, to be understood here as the experimental reference temperature. In general, $\tau_{\omega}= \tau (\omega, T)$; however, as a result of the linearization, $\tau_{\omega}= \tau (\omega, T_\text{eq}) \equiv \tau (\omega)$; in other words, the solutions (and associated reconstruction) are valid for the experiment baseline temperature $T_\text{eq}$. Also, $(de^\text{eq}/dT)_{T_\text{eq}}= \hbar \omega ( df^\text{eq}_T/dT ) \vert_{T_\text{eq}}$ and $f^\text{eq}_{T}$ is the Bose-Einstein distribution with temperature parameter $T$, given by
\begin{equation} \label{BE}
f^\text{eq}_{T} (\omega) = \frac{1} {\exp (\hbar \omega/ k_B T)- 1} ,
\end{equation} 
where $k_B$ is Boltzmann's constant. Finally, $\Delta \widetilde{T}= \Delta \widetilde{T} (t, \mathbf{x})= \widetilde{T}- T_\text{eq}$ is referred to as the deviational pseudo-temperature ($\widetilde{T} (t, \mathbf{x} )$ is the pseudo-temperature). Note that the deviational pseudo-temperature, which is different from the deviational temperature defined below, is defined by the energy conservation statement \cite{ARHT}
\begin{equation}
\int_{\mathbf{\Omega}} \int_{\omega} \left[ \frac{C_\omega} {\tau_{\omega}} \Delta \widetilde{T} - \frac{e^d} {\tau_{\omega}} D_{\omega} \right] d \omega d \mathbf{\Omega}= 0 , \label{pseudotemperature}
\end{equation} 
in which $D_{\omega}= D(\omega)$ is the density of states, $C_{\omega}= C (\omega; T_\text{eq})= D_{\omega} (de^\text{eq}/dT)_{T_\text{eq}}$ is the frequency-dependent volumetric heat capacity, and $d \mathbf{\Omega}= \sin (\theta) d \theta d \phi$ represents the differential solid angle element such that $\theta$ and $\phi$ refer to the polar and azimuthal angles in the spherical coordinate system, respectively. The temperature $T (t, \mathbf{x})$ is computed from 
\begin{equation}
\int_{\mathbf{\Omega}} \int_{\omega} \left[ C_{\omega} \Delta T- e^d D_{\omega} \right] d \omega d \mathbf{\Omega}= 0 , \label{temperature}
\end{equation}
where $\Delta T (t, \mathbf{x})= T- T_\text{eq}$ is the deviational temperature. The frequency-dependent free path is given by 
\begin{equation} 
\Lambda_{\omega}= v_{\omega} \tau_{\omega} ,
\label{lambda}
\end{equation} 
where $v_{\omega}= || \mathbf{v}_{\omega} ||$ is the group velocity magnitude. Cumulative distribution function (CDF) of thermal conductivity as a function of the free path, introduced in section \ref{intro}, is defined as $F (\Lambda):= \frac{1} {3 \kappa} \int_{\omega^*(\Lambda)} C_{\omega} v_{\omega}^2 \tau_{\omega} d \omega$, where $\omega^*(\Lambda)$ is the set of modes such that $\omega^*(\Lambda)= \{\omega | \Lambda_{\omega} \leq \Lambda\}$; similarly, the corresponding probability density function (of thermal conductivity) is given by $\mathfrak{f}= \frac{dF}{d\Lambda}$.

\section{Limitations associated with $F(\Lambda)$}
\label{free_path_material}

In this section, we investigate, both numerically and analytically, whether $F(\Lambda)$ can be used to describe heat transfer in all transport regimes. This investigation is motivated by the widespread use of $F(\Lambda)$ in procedures for reconstructing material properties from thermal spectroscopy experiments. 

The approximations and theoretical inconsistencies associated with the use of the concept of effective thermal conductivity, in conjunction with $F(\Lambda)$, to reconstruct thermal spectroscopy data have been discussed before in \cite{Mojtaba2016, Mojtaba2018}, as well as section \ref{intro}. Here, we take a different approach which allows us to examine the premise of such approaches on a more general level.  First, we prove through a theorem that a given $F(\Lambda)$ may arise as a result of more than one distribution of relaxation-times. Since different relaxation-time distributions will, in general, lead to different material thermal behaviors, this implies that $F(\Lambda)$ cannot uniquely determine the material thermal behavior, which in turn questions the ability of $F(\Lambda)$ to describe thermal responses in sub-micron regimes. The above assertion is also demonstrated through a numerical experiment using a realistic material model for silicon.
 
\paragraph{Theorem:} For a given material, if there are two phonon frequencies, $\omega_1$ and $\omega_2$, such that $C_\omega ({\omega_1}) v_\omega({\omega_1})= C_\omega({\omega_2}) v_\omega({\omega_2})$, there exist more than one relaxation-time distributions that lead to a given $F(\Lambda)$. 

The proof of theorem is left to Appendix A.

\subsection{Numerical Demonstration} 
\label{numer_dem}

In order to highlight the importance of the above in practical terms, namely, the existence of multiple relaxation-time functions for a given $F(\Lambda)$, and more importantly, the non-uniqueness of the thermal response for a given $F(\Lambda)$, we have performed a calculation that directly studies this hypothesis. In the interest of simplicity, all calculations were performed for the 1D-TTG geometry with the Holland model for material Si~\cite{Mojtaba2016, Mojtaba2018}. Readers are referred to Appendix B for a discussion of the functional form of the product $C_\omega ({\omega}) v_\omega({\omega})$ for Si, used in our calculations. We consider the objective function 
\begin{equation}
\min_{{\bf U}} \int_{\Lambda} \big| \mathfrak{f}^*(\Lambda)- \mathfrak{f}(\Lambda) \big| d\Lambda ,
\label{optim_unique}
\end{equation}
where $\mathfrak{f}^*$ represents the ``true" free path-dependent distribution function of thermal conductivity and ${\bf U}$ is a set of parameters that parameterizes the relaxation-times function. Here, without loss of generality, we have assumed ${\bf U}$ to be a set of parameters that allows different branches of the relaxation-times function to take the form of a piece-wise linear function with three segments as a function of the phonon frequency; this is the same parameterization that has been used previously in \cite{Mojtaba2016, Mojtaba2018, MojtabaAPL}. This objective function attempts to find free path-dependent thermal conductivity distribution functions $\mathfrak{f}(\Lambda)$ that match the true thermal conductivity distribution function by searching in the parameter space ${\bf U}$. Note that ${\bf U}$ enters the objective function through the definition of the free path distribution $\Lambda_\omega=v_\omega \tau_\omega({\bf U})$. 

\begin{figure}[htbp]
\centering
\begin{subfigure}{.32\textwidth}
\centering
\includegraphics[width=1.1\linewidth]{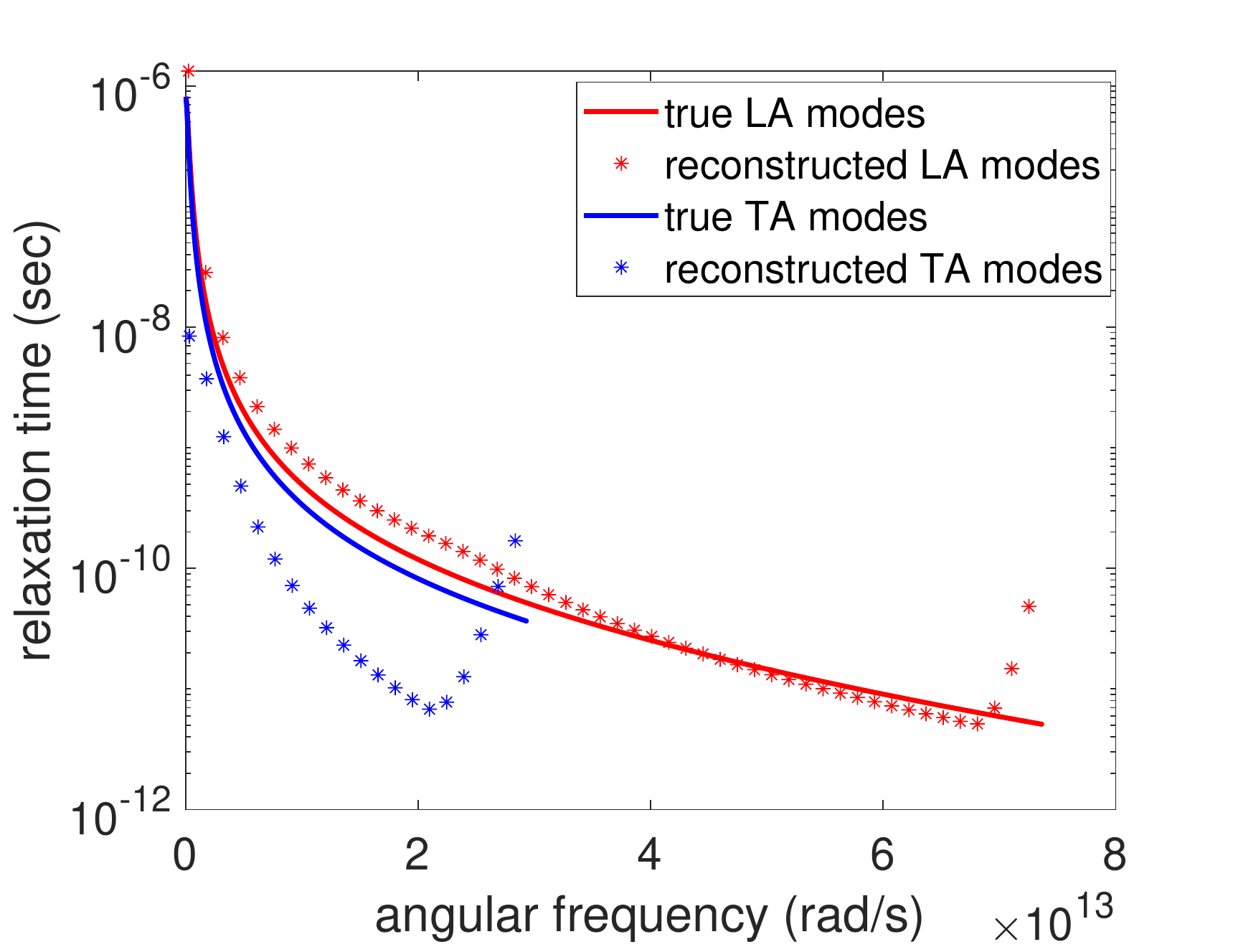}
\caption{Relaxation-time}
\label{ab_initiomodel}
\end{subfigure}
\begin{subfigure}{.32\textwidth}
\centering
\includegraphics[width=1.1\linewidth]{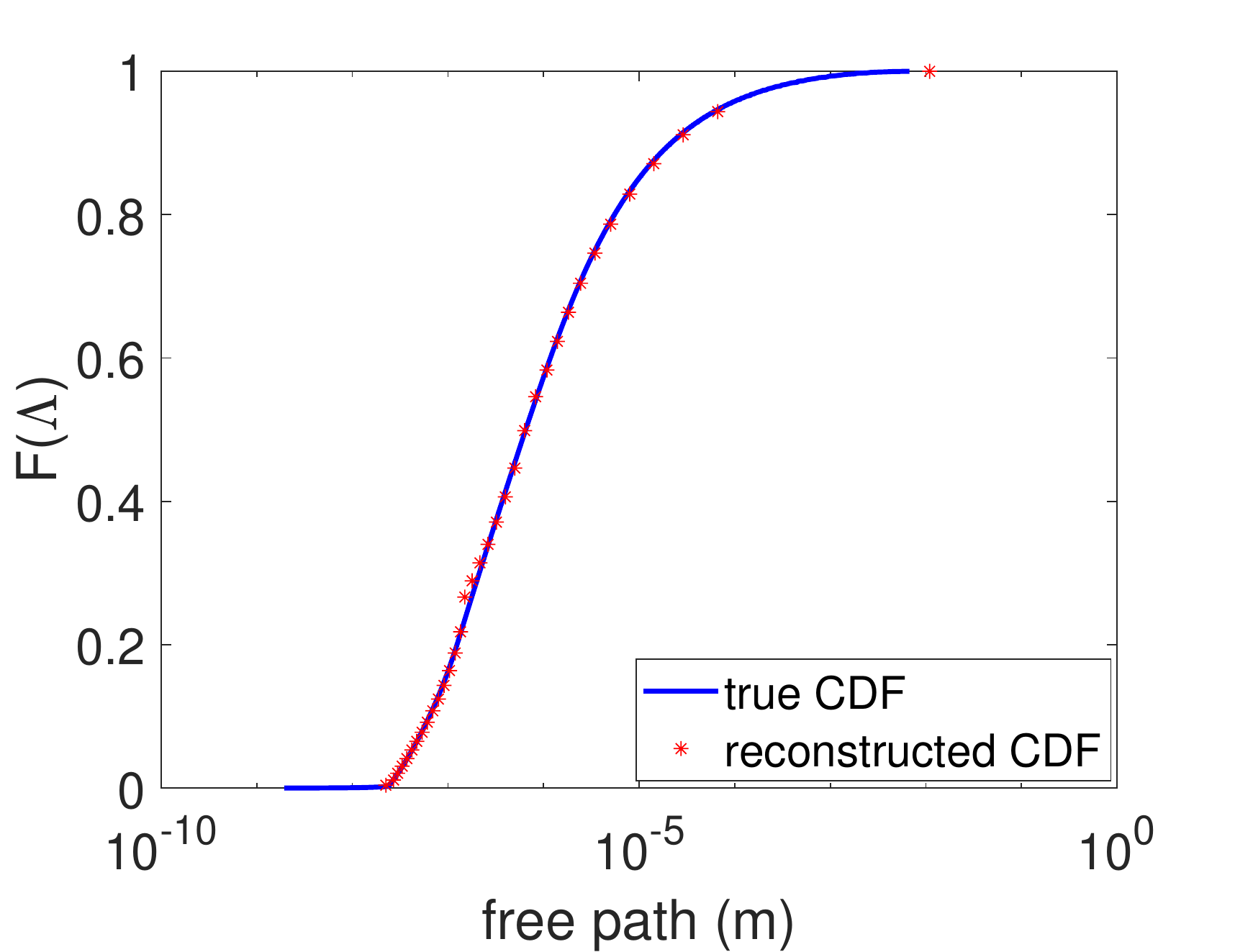}
\caption{$F(\Lambda)$}
\label{Hollandmodel}
\end{subfigure}
\begin{subfigure}{.32\textwidth}
\centering
\includegraphics[width=1.1\linewidth]{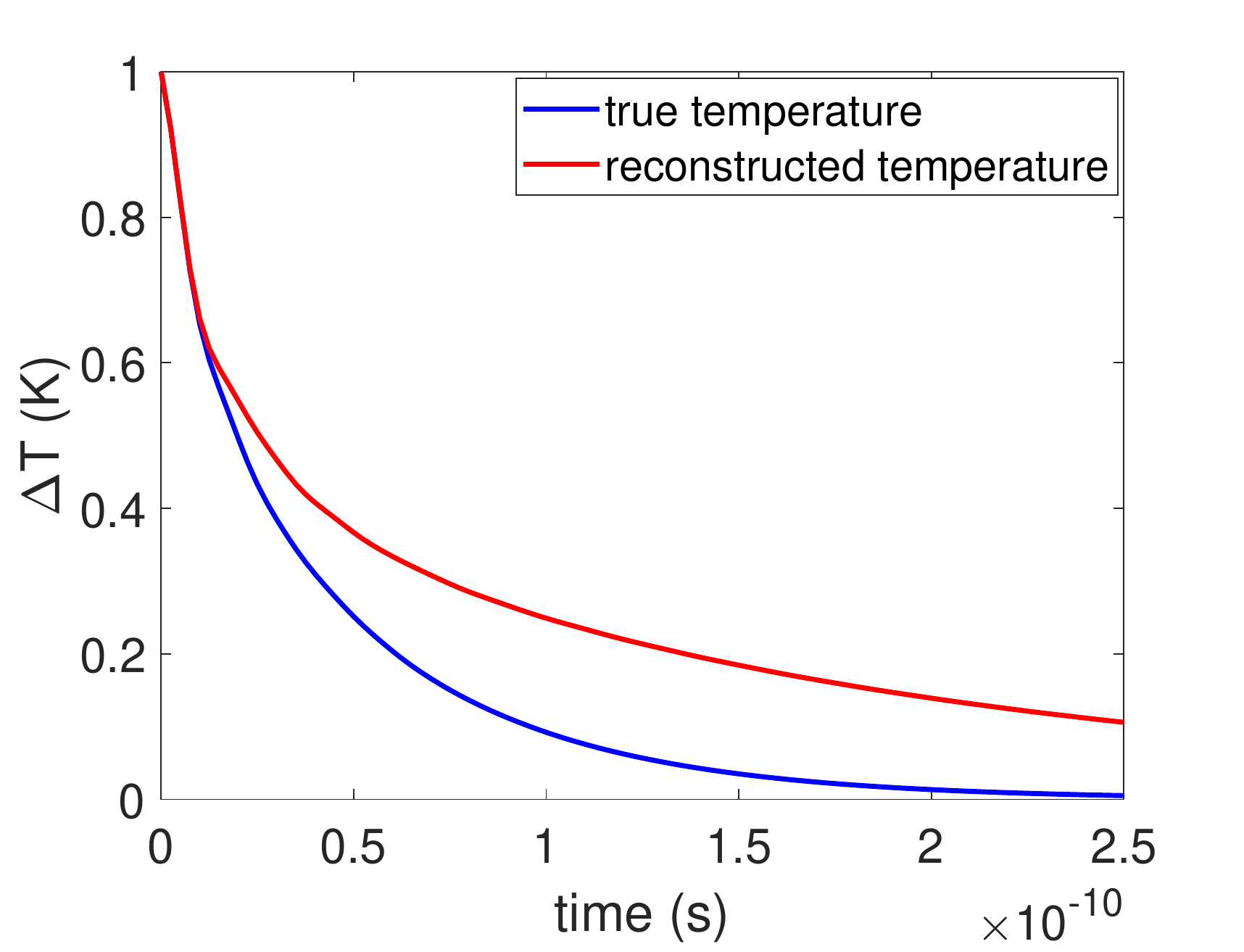}
\caption{Temperature}
\label{T_100n_unique}
\end{subfigure}
\caption{Comparison between properties predicted by one representative solution of optimization \eqref{optim_unique} and true data. As expected, although the two $F(\Lambda)$ match, the relaxation-times and thermal responses are clearly different.}
\label{k_unique}
\end{figure}

We perform the optimization using the Nelder-Mead (NM) algorithm \cite{NelderMead}, repeating it multiple times starting from different initial conditions for ${\bf U}$ in order to also study the effect of different initial conditions on the optimization process. Figure \ref{k_unique} shows the relaxation-times and the $F(\Lambda)$ obtained from one of these optimization processes as well as the temperature profile predicted by the optimized ${\bf U}$ versus the true profile at 100 nm length scale. This profile is generated using the inverse fast Fourier transform (IFFT) \cite{IFFT_book} as the forward simulation method \cite{Mojtaba2016}. As it is expected from equation \eqref{optim_unique}, the reconstructed $F(\Lambda)$ matches the true CDF. At the same time, both the relaxation-time functions, and more importantly, the thermal responses do not match, even though all other material properties, as well as the geometry are kept fixed. The mismatch in thermal responses also exists at other length scales.

The results provided in figure \ref{k_unique} clearly show, from a numerical perspective, that $F(\Lambda)$ cannot be used reliably as a material property for predicting thermal behaviors in all regimes. While the non-uniqueness of the relaxation-times function for a given $F(\Lambda)$, as was proven theoretically in this section and shown numerically in figure \ref{ab_initiomodel}, may not be catastrophic on its own, the fact that these different functions predict different thermal responses, as we have seen in figure \ref{T_100n_unique}, implies that a given $F(\Lambda)$ provides insufficient information for predicting the thermal response. The results provided in this section prove that {\bf $F(\Lambda)$ cannot be used to predict thermal responses in all transport regimes}.

\section{An alternative to $F(\Lambda)$}
\label{Unique_analysis}
The discussions provided previously in \cite{Mojtaba2016, Mojtaba2018, MojtabaAPL}, and more importantly, in the previous section, show that $F(\Lambda)$, regardless of the method being used to reconstruct it and its accuracy, is not able to predict the material thermal behavior in all regimes, and the  sub-micron regime in particular, which is of interest presently. This is in sharp contrast with the relaxation-times function which is related to the system thermal response through a direct, one-to-one relationship. 

In previous work we proposed a multi-stage NM algorithm for reconstruction of relaxation-times function from thermal spectroscopy experiments \cite{Mojtaba2016, Mojtaba2018, MojtabaAPL} and validated it both numerically and experimentally. In this section, we briefly study the reliability of the solutions of this previously proposed optimization algorithm, that is, the uniqueness of the optimization solution as well as a Bayesian-approach inspired study of the sensitivity of the solutions to measurement noise. Due to the complexity of the BTE and the related inverse problems, here we have focused on the numerical study of a specific problem, the 1D-TTG experiment.

Due to the success of the NM algorithm in providing solutions to the reconstruction problem \cite{Mojtaba2016, Mojtaba2018, MojtabaAPL}, here, we use a globalized version of NM algorithm, referred to as the ``globalized bounded Nelder-Mead (GBNM)" \cite{GNM}, to study the uniqueness of the solutions to the reconstruction problem formulation used in \cite{Mojtaba2016, Mojtaba2018, MojtabaAPL}. The GBNM algorithm is very similar to the NM algorithm described before; however, instead of performing the optimization once, starting from one initial condition, it repeats the complete optimization process many times, starting from different probabilistically correlated initial conditions. More details can be found in \cite{GNM}.

\subsection{On the uniqueness of reconstructed solutions}
To assess the uniqueness of the reconstructed solutions obtained via the optimization formulation, we follow a procedure that is similar to our previous work \cite{Mojtaba2016}: we generate synthetic temperature profiles for a material by solving the BTE, and use the generated data to infer the material relaxation-time by solving the following optimization problem 
\begin{equation}
\min_{{\bf U}} \mathcal{L}= \min_{{\bf U}} \left[ \frac{ \sum_{t, \textbf{x}, L} | T_\text{m} (t, \mathbf{x}; L)- T_\text{BTE} (t, \mathbf{x}; L, {\bf U}) | } {N}+ \alpha \Bigg| 1- \frac{1} {3\kappa} \int_{\omega} C_{\omega} \tau_{\omega} ({\bf U}) v^2_{\omega} d \omega \Bigg| \right] , \label{objectivefunction_2TA} 
\end{equation}
where $T_\text{m} (t, \mathbf{x}; L)$ denotes the experimentally measured temperature (equivalent to synthetically generated temperatures), $T_\text{BTE}$ is the temperature computed from solution of the BTE (the same temperature as in equation \eqref{temperature}), $N$ is the total number of (independent) measurements available, and $L$ is the different characteristic length scales. In the present case, we have assumed the material to be Si and considered its response in the 1D-TTG geometry for 10 nm$<L<$100 $\mu$m, solved for using the IFFT method \cite{IFFT_book, Mojtaba2016}. We also assume that the two $TA$ branches of the material are the same, using the same piece-wise linear parameterization of \cite{Mojtaba2016}, for both the \textit{ab initio} and the Holland models~\cite{Mojtaba2016, Mojtaba2018, MojtabaAPL}.

\begin{figure}[htbp]
\centering
\begin{subfigure}{.49\textwidth}
\centering
\includegraphics[width=1.0\linewidth]{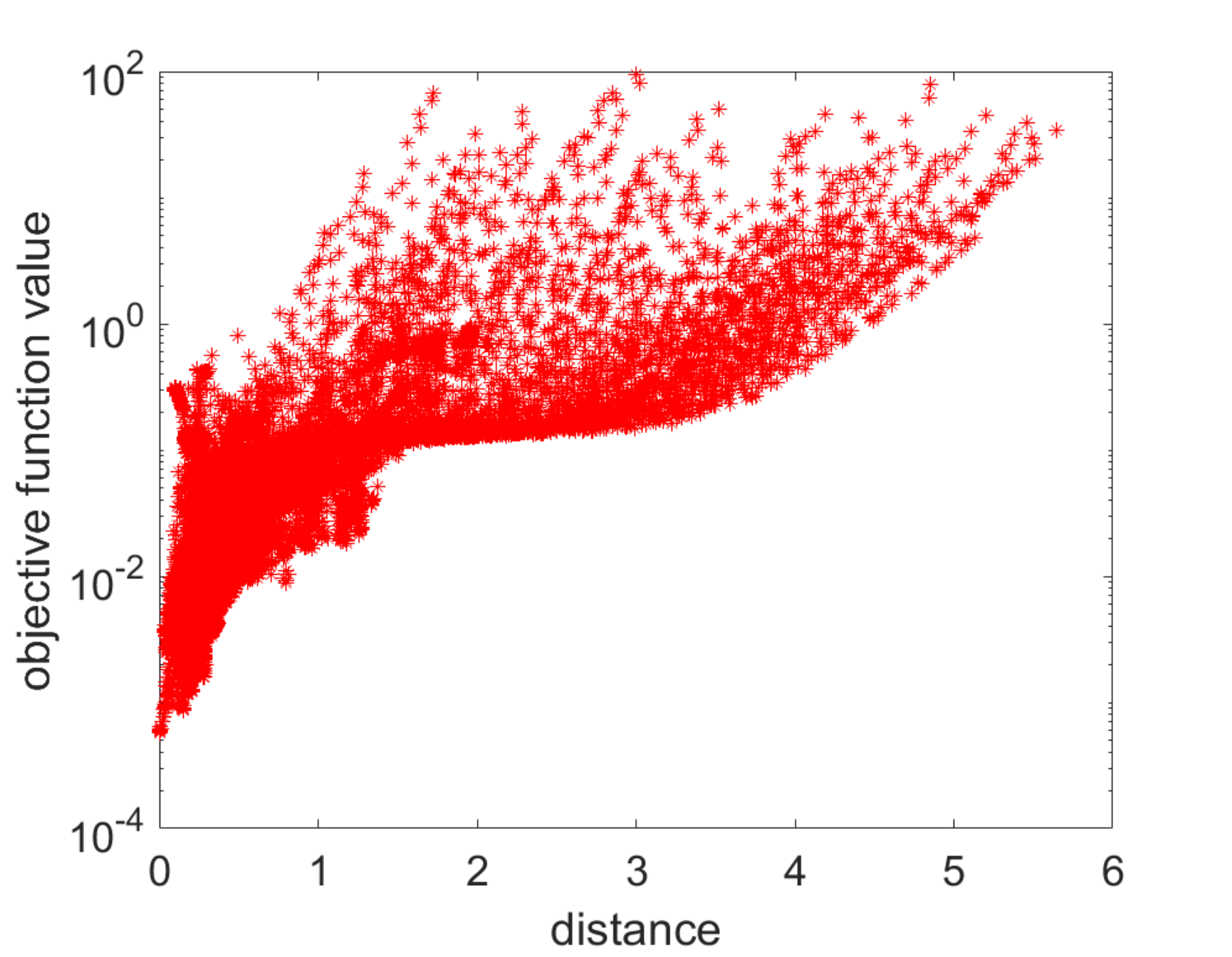}
\caption{}
\label{ab_initiomodelz}
\end{subfigure}
\begin{subfigure}{.49\textwidth}
\centering
\includegraphics[width=1.0\linewidth]{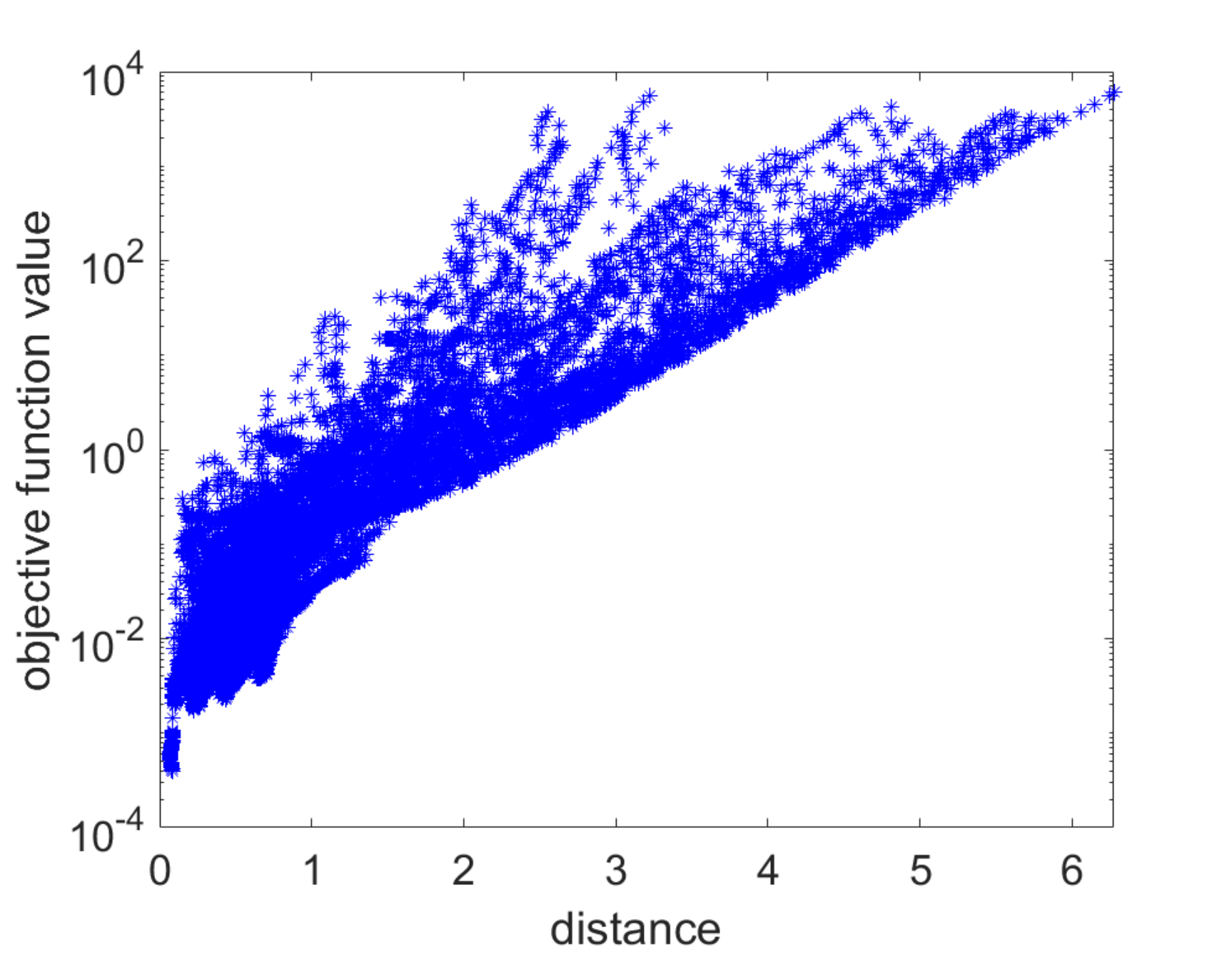}
\caption{}
\label{Hollandmodel}
\end{subfigure}
\caption{The value of the objective function versus the distance (see equation \eqref{distance}) between the points sampled during GBNM and the true solution for the Holland model (a) and the \textit{ab initio} model (b).}
\label{GNM_plot}
\end{figure}

As stated above, the difference from previous work is that, here, the optimization is performed using the GBNM method. Figure \ref{GNM_plot} shows the value of the objective function for the various points that the GBNM method explored on its way to minimizing ${\cal L}$. More detail on the procedure and parameters used here can be found in \cite{MojtabaThesis}. The horizontal axis in the figure corresponds to the distance between a ``trial" relaxation-times function (whose value of ${\cal L}$ is plotted on the ordinate) and the ``true" solution $\tau^*_\omega$; specifically, ``distance" is defined by
\begin{equation}
\text{distance}:= \frac{{\bigintss}_{\omega} \left| \log\left( \frac{\tau_\omega}{\tau^*_{\omega}} \right) \right| D_\omega d\omega }{\int_{\omega} D_\omega d\omega}.
\label{distance}
\end{equation}
Here, $\omega$, as before, contains information of both different frequencies and branches. 

The main message from figure \ref{GNM_plot} is that the bottom left of the cluster of datapoints in both plots represents a monotonically increasing function, which means that the closer we are to the true solution, the smaller the value of the objective function becomes, and more importantly, this is a monotonic behavior as we approach the true solution. This implies that the solution is unique; non-uniqueness would be implied by points in the lower right hand corner of the diagrams (low value of objective function, while the distance from the solution is large). Here we note that the number of datapoints in each plot in figure \ref{GNM_plot} is approximately 1 million. While exploring the whole parameter space with high resolution is not feasible due to the high dimension of the problem studied here ($\text{dim}({\bf U})=12$; two branches, each parameterized with three lines), the general behavior of the objective function, such as the monotonicity described above, will likely not change if more iterations are used (and consequently an even better exploration of the parameter space is performed). 

\subsection{Sensitivity of reconstructed relaxation-time to measurement uncertainty}
While the results provided in figure \ref{GNM_plot} show that the behavior of the objective function $\cal{L}$ introduced in \cite{Mojtaba2016} resembles that of a function with a unique minimum, the sensitivity of this solution to noise in the measurement is not clear. Low sensitivity of an objective function to noise plays an important role in its practical application to real-world  measurements. Here we study this sensitivity through the Bayesian theorem, that is, using the relation
\begin{equation}
p\Big( {\bf U} \big| \{ T_\text{m}(t, {\bf x}=0; L)\}_{t,L}\Big) \propto p\Big(\{ T_\text{m}(t, {\bf x}=0; L)\}_{t,L} \big| {\bf U}\Big) \pi({\bf U}) ,
\label{Bayes}
\end{equation}
where $p\Big({\bf U} \big| \{ T_\text{m}(t, {\bf x}=0; L)\}_{t,L}\Big)$ is the distribution of the parameters of the relaxation-times function given the temperature measurements (the posterior distribution) --- the quantity that we are interested in--- from which we can also calculate the distribution of the relaxation-times function itself through the $\tau_\omega({\bf U})$ function; $p\Big(\{ T_\text{m}(t, {\bf x}=0; L)\}_{t,L} \big| {\bf U}\Big)$ is the distribution of measured temperatures, given the parameters of the relaxation-times function (the likelihood function). The likelihood function can be calculated by first solving the BTE for a given relaxation-times function $\tau_\omega({\bf U})$, and then adding an artificial noise to its predicted temperatures that resembles that of noisy measurements. Finally, $\pi({\bf U})$ is the prior distribution which is usually a wide (least informative) distribution of the quantity of interest, the relaxation-times function; for instance, a wide symmetrical distribution around the correct relaxation-times function. More detailed definitions and information on equation~\eqref{Bayes} are provided in Appendix C. 

\begin{figure} [htbp]
\begin{subfigure}{.5\textwidth}
\centering
\includegraphics[width=1.0\linewidth]{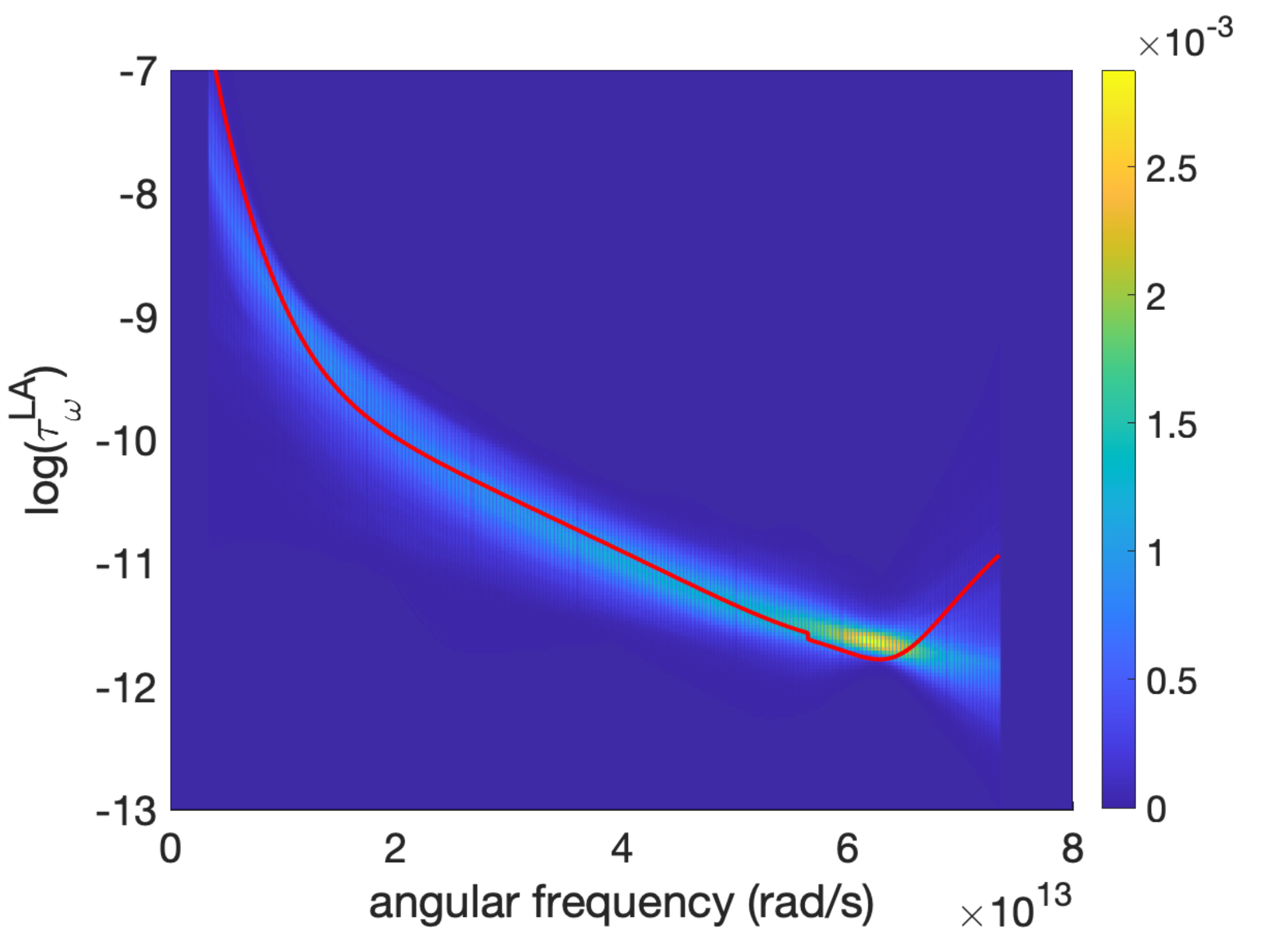}
\caption{$LA$ modes}
\label{NM1}
\end{subfigure}
\begin{subfigure}{.5\textwidth}
\centering
\includegraphics[width=1.0\linewidth]{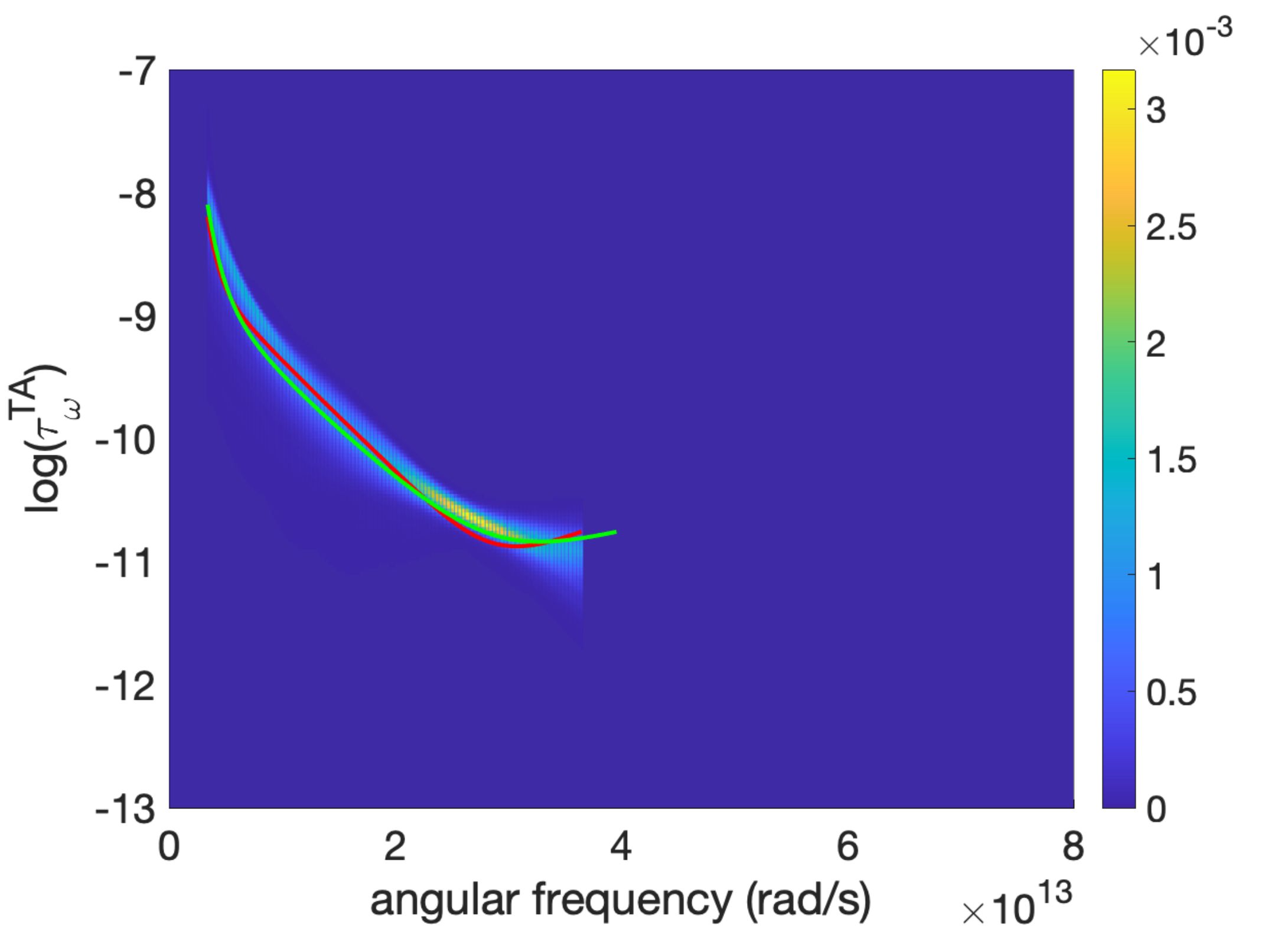}
\caption{$TA$ modes}
\label{NM2}
\end{subfigure}
\caption{Contour plot of the probability density of the frequency-dependent relaxation-times distribution of $LA$ and $TA$ modes. The red line in (a) denotes the true $\tau_\omega^{LA}$, while the red line in (b) denotes the true $\tau_\omega^{TA_1}$ and the green line in (b) denotes the true $\tau_\omega^{TA_2}$. }
\label{LA_TA_bayes}
\end{figure}

Once we infer the distribution of the different components of {\bf U} (the posterior distribution), we can calculate the distribution of the relaxation-times function as a function of its frequency $\tau_\omega({\bf U})$. The results for the {\it ab initio} silicon model are provided in figure \ref{LA_TA_bayes}. We have also plotted the true relaxation-time functions in this figure. We notice that the distribution matches the true solution ``in the mean''. We also observe that for $LA$ modes, at frequencies around $\omega= 6\times 10^{13}$ rad/s, the distribution is sharper than other frequencies, implying that a relatively more accurate reconstruction of the value of relaxation-time at those frequencies is possible. On the other hand, the uncertainty at very large frequencies ($\omega \gtrsim 6.5\times 10^{13}$ rad/s) is more than the other ranges of the frequency. A similar trend can be observed for $TA$ modes. Overall, by comparing the two plots in figure \ref{LA_TA_bayes}, we conclude that the reconstruction of $TA$ branches is more accurate (the distributions are sharper). This could be due to the assumption that the two $TA$ branches have been approximated by the same function, which consequently led to their greater influence on the thermal behavior.

\begin{figure}[htbp]
\begin{subfigure}{.5\textwidth}
\centering
\includegraphics[width=1.0\linewidth]{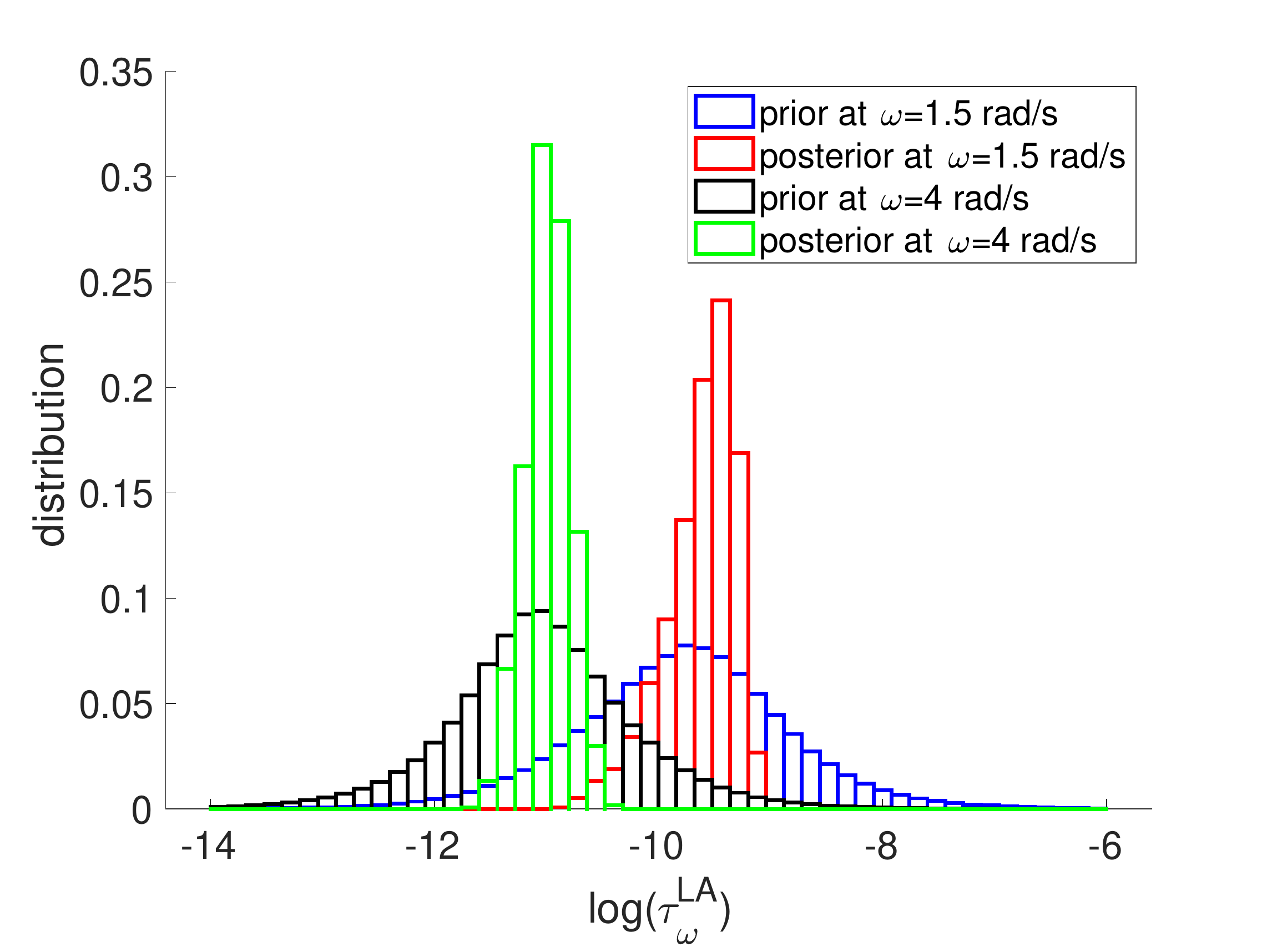}
\caption{}
\label{NM1}
\end{subfigure}
\begin{subfigure}{.5\textwidth}
\centering
\includegraphics[width=1.0\linewidth]{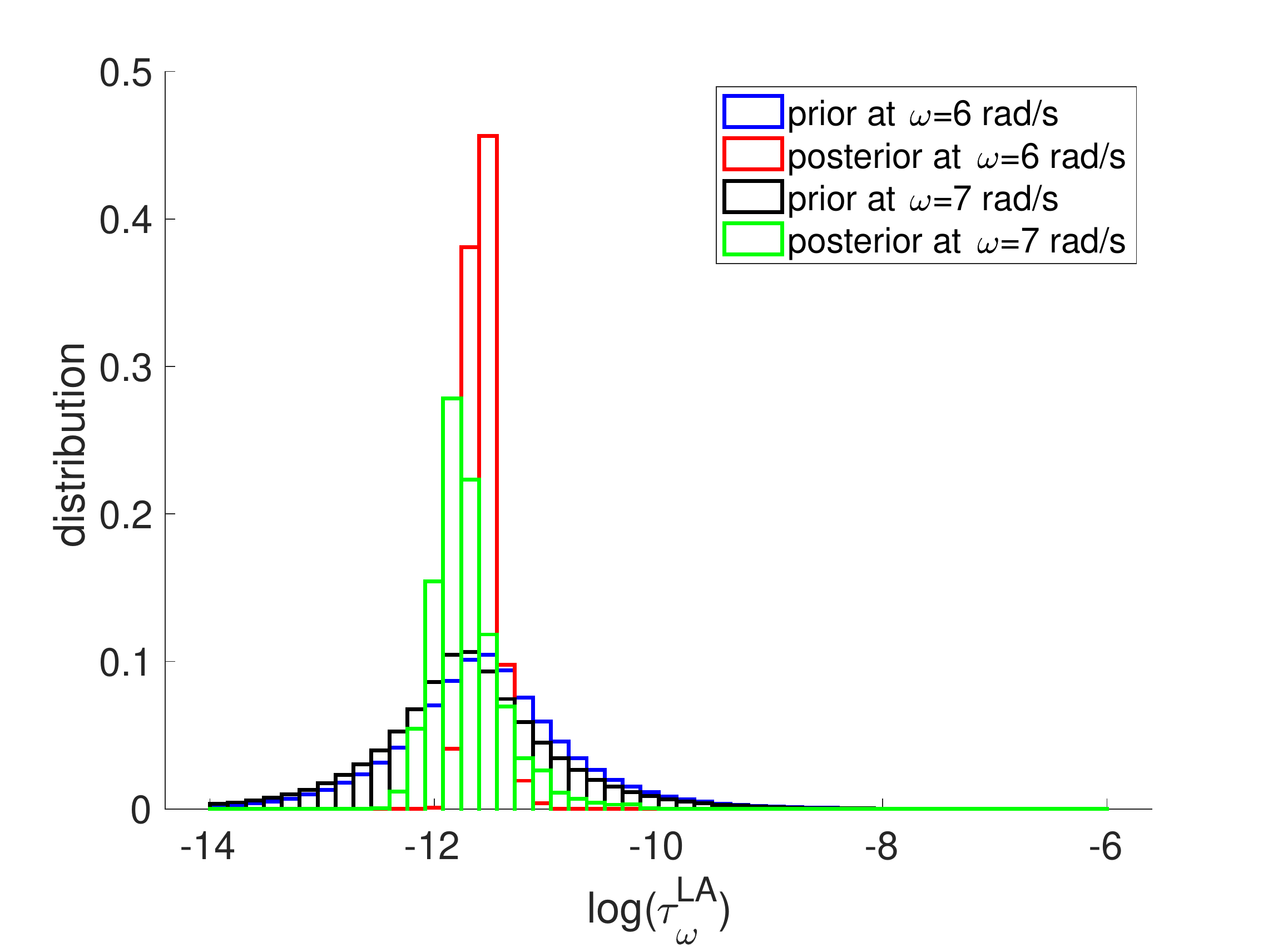}
\caption{}
\label{NM1}
\end{subfigure}
\caption{Distributions at a few low frequency (a) and high frequency (b) $LA$ modes. }
\label{LA_bayes_prob}
\end{figure}

\begin{figure} [htbp]
\begin{subfigure}{.5\textwidth}
\centering
\includegraphics[width=1.0\linewidth]{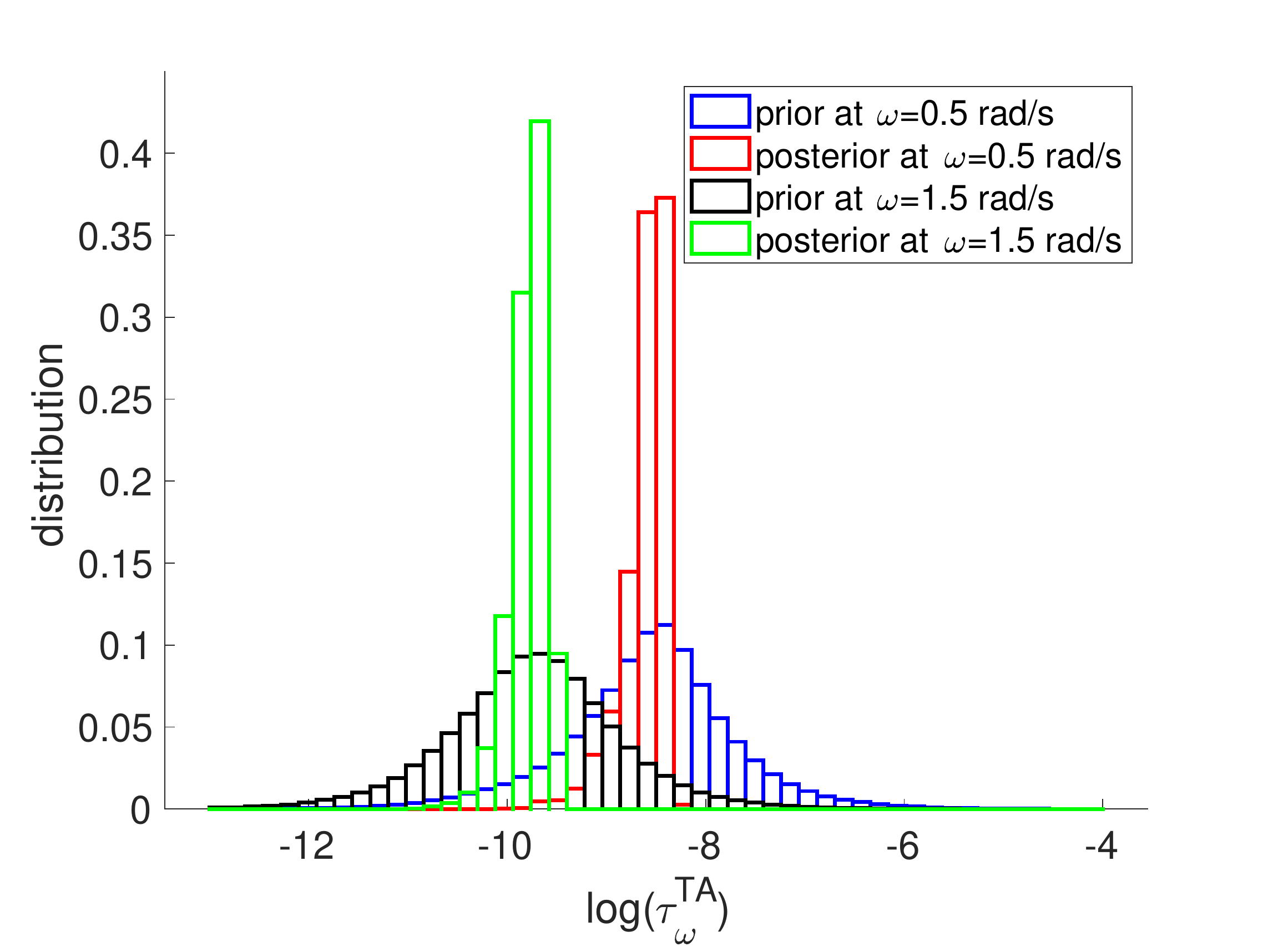}
\caption{}
\label{NM1}
\end{subfigure}
\begin{subfigure}{.5\textwidth}
\centering
\includegraphics[width=1.0\linewidth]{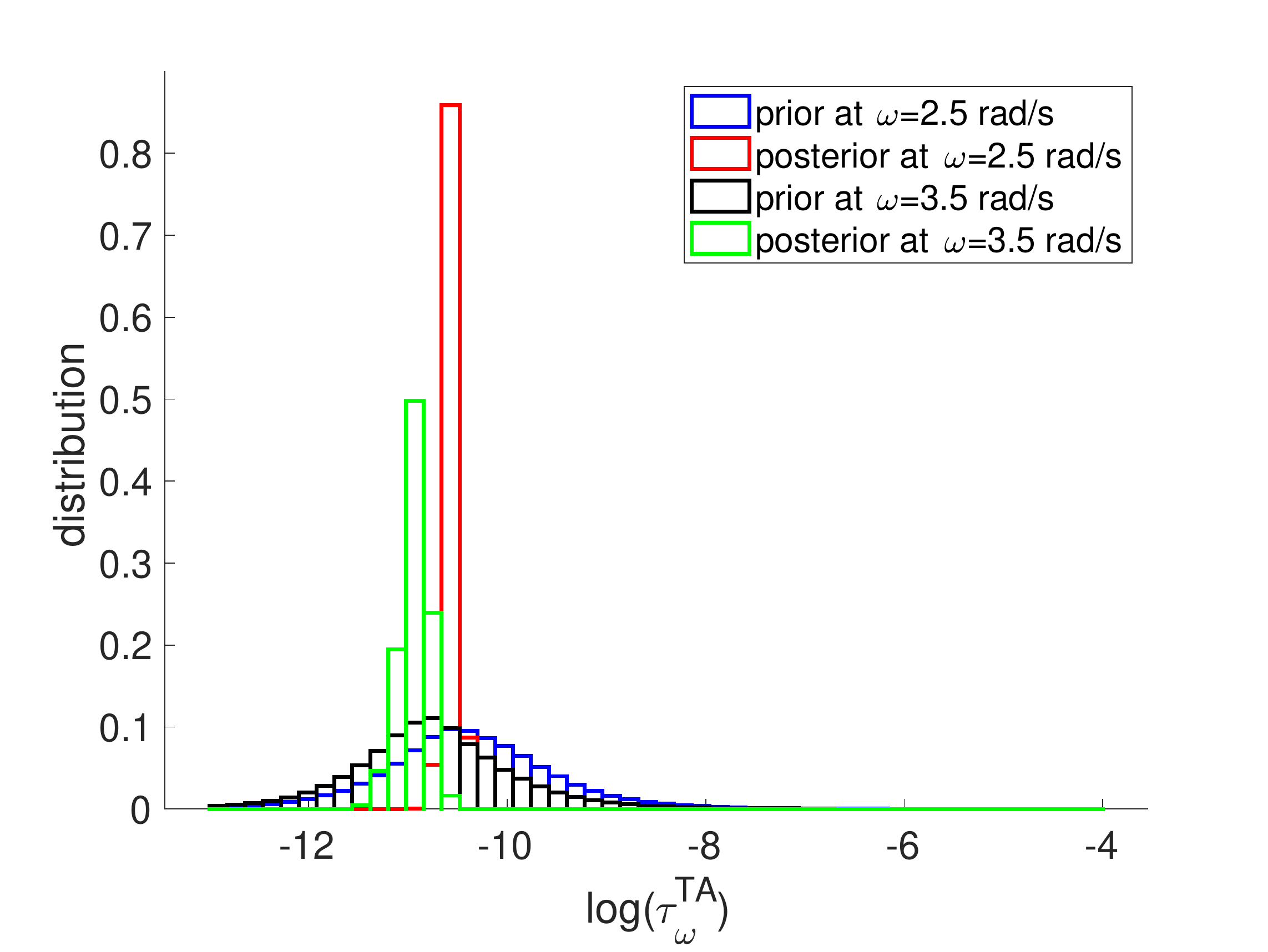}
\caption{}
\label{NM1}
\end{subfigure}
\caption{Distributions at a few low frequency (a) and high frequency (b) $TA$ modes. }
\label{TA_bayes_prob}
\end{figure}

In order to obtain a better picture of the shape of the distributions at different frequencies, we have also plotted the distributions at a few specific frequencies (equivalent to cross-sections of figure \ref{LA_TA_bayes} at different frequencies). Figures \ref{LA_bayes_prob} and \ref{TA_bayes_prob} show the prior distribution, $\pi(\tau_\omega)$, versus the posterior distribution, $p(\tau_\omega \big| \{ T_\text{m}(t, {\bf x}=0; L)\}_{t,L})$, for a few frequencies of $LA$ and $TA$ modes, respectively. We observe that in most cases, the posterior distribution has  become significantly narrower compared to the prior distribution, implying that the chosen prior distribution has been wide enough to avoid bias in inferring the posterior distribution. The distributions in the $TA$ case are sharper, as can also be seen in figure \ref{LA_TA_bayes}. The distribution at very high frequencies ($\omega= 7$ rad/s in figure  \ref{LA_bayes_prob} and $\omega= 3.5$ rad/s in figure  \ref{TA_bayes_prob}) is wider than the other frequencies, consistent with our previous observations from figure \ref{LA_TA_bayes}. While the results provided in figure \ref{GNM_plot} point to the uniqueness of the objective function minimum point, the narrow distributions observed in figures \ref{LA_TA_bayes}--\ref{TA_bayes_prob} complement those results by implying that good accuracy of solution in the vicinity of the optimal solution, minimally affected by the measurement noise, is attainable.  

\section{Summary and outlook} 
\label{Conclude}
We have studied the type and reliability of information that can be extracted from thermal spectroscopy experiments. We specifically showed that inverting the experimental data using $F(\Lambda)$ (presumably via the effective thermal conductivity construct) leads to an ill-posed problem because any given $F(\Lambda)$ {\it  does not} correspond to a unique system thermal response; as a result, even correct determination of $F(\Lambda)$ from the experiment is insufficient by itself to fully characterize the system behavior in other situations. Although the condition $C_\omega ({\omega_1}) v_\omega({\omega_1})= C_\omega({\omega_2}) v_\omega({\omega_2})$ in section \ref{free_path_material} is more general, in practice, this condition is typically met due to the presence of more than one phonon branch, which $F(\Lambda)$ is unable to distinguish between. Our previous results in \cite{Mojtaba2016, Mojtaba2018, MojtabaAPL, MojtabaThesis}---also briefly reviewed in section~\ref{intro}--- additionally showed that the Fourier heat conduction equation with effective thermal conductivity (or diffusivity) can be used for the reconstruction of $F(\Lambda)$ only if certain late time and large scale assumptions are met (see \cite{Mojtaba2016, Mojtaba2018} and figures~\ref{k_eff_issue} and \ref{k_eff_issue_2} of the present manuscript), and is thus not applicable at arbitrarily small length scales.

Our results in section \ref{Unique_analysis} are consistent with our previous work, showing that the algorithm proposed in \cite{Mojtaba2016} can be reliably used for future reconstruction purposes on real experimental data (as it has already been performed in \cite{MojtabaAPL}) which may include noise in their measurements, with no concerns about the non-uniqueness of the solution or low sensitivity of the relaxation-times function to the measured thermal responses. Note that although finding this unique solution requires a global optimization algorithm which is significantly more expensive, possibly by orders of magnitude, than the multi-stage NM optimization algorithm used in \cite{Mojtaba2016,Mojtaba2018,MojtabaAPL}, this is not required for regular applications.  In applications, the multi-stage algorithm proposed in \cite{Mojtaba2016} should provide a solution that is sufficiently close to the global optimum at a fraction of the cost. Specifically, 1 million forward simulations (the number of datapoints shown in figure \ref{GNM_plot}) were used to find the approximate location of the global minimum using GBNM, while as it was discussed in \cite{Mojtaba2016}, our multi-stage algorithm requires at most about 1000 forward simulations to obtain this solution.

While in the present work the numerical analysis of the reliability and uniqueness of our previously proposed algorithm~\cite{Mojtaba2016} has been focused on the 1D-TTG experiment, it is important to note that many thermal spectroscopy experiments also include solid-solid interfaces in their setup, which can add to the complexity of the analysis, for instance, due to lack of sufficient information regarding the interface conductance and transmissivities. However, in a previous work~\cite{Mojtaba2018}, we have studied the reliability of our proposed reconstruction process on the 2D-dots geometry~\cite{Lingping_nature} and have observed that not knowing the interface transmissivities does not influence the reconstruction process significantly, and, in fact, the phonon relaxation-times function can be reconstructed accurately without knowing any information of the transmissivity functions. In future, we will also perform a study similar to present work that aims at uniqueness and sensitivity of the reconstructed relaxation-times in the presence of solid-solid interface with unknown properties.

The processes being described in section \ref{Unique_analysis} introduce new complementary tools for analyses of thermal spectroscopy measurements and open new avenues for future similar studies by combining them with the available theoretical tools, leading to a unified numerical-analytical approach for categorization of different problems based on the uniqueness of their inverse problems, in the context of phonon transport. This will allow researchers to explore the uniqueness of the solution for a particular geometry before performing the experiment, and thus helping them to focus on experiments whose solution to their inverse BTE equations is well-posed. Similarly, the Bayesian approach can help with visualization of sensitivity of the solution around the true solution, thus guiding the researchers towards geometries that reveal high sensitivity in the vicinity of the solution, and consequently, are more robust to measurement noise.

\section*{Acknowledgment}
The authors would like to thank Y. M. Marzouk and G. Chen for their comments and suggestions. This work was supported by the Solid-State Solar-Thermal Energy Conversion Center (S3TEC), an Energy Frontier Research Center funded by the U.S. Department of Energy, Office of Science, Basic Energy Sciences, under Award\# DE-SC0001299 and DE-FG02-09ER46577.

\appendix

\section*{Appendix A: Proof of theorem}
{\it Proof:} We consider two different relaxation-time functions, $\tau_\omega$ and $\tau'_\omega$, with free path-dependent distribution function of thermal conductivity $\mathfrak{f}(\Lambda)$ and $\mathfrak{f}'(\Lambda)$, respectively; $\tau_\omega$ and $\tau'_\omega$ are identical ($\tau_\omega=\tau'_\omega$) for all $\omega\neq \omega_1,\omega_2 $, while at $\omega_1$, $\omega_2$ they satisfy the relations 
\begin{equation}
\tau'_\omega(\omega_1)= \frac{\tau_\omega(\omega_2) v_\omega(\omega_2)}{ v_\omega(\omega_1)} \ \text{,}\ \tau'_\omega(\omega_2)= \frac{\tau_\omega(\omega_1) v_\omega(\omega_1)}{ v_\omega(\omega_2)} ,
\label{taus_rel}
\end{equation}
where we assume $v_\omega(\omega_2)\neq v_\omega(\omega_1)$.

From the relaxation-times function $\tau_\omega$ one can derive the free path $\Lambda_\omega$ and frequency-dependent heat conductivity $\kappa_\omega$ ($\kappa_\omega := C_\omega v_\omega^2 \tau_\omega/3$), considered here as a pair $(\Lambda_\omega,\kappa_\omega)=(v_\omega\tau_\omega, C_\omega v_\omega^2 \tau_\omega/3)$. Since the two relaxation-times functions are identical at all $\omega\neq \omega_1,\omega_2$, their corresponding $(\Lambda_\omega,\kappa_\omega)$ pairs are also identical at all $\omega\neq \omega_1,\omega_2$. To show that their pairs are identical for all frequencies, and thus {\it they have the same CDF of thermal conductivity}, we now show that
\begin{equation}
(\Lambda_\omega(\omega_1),\kappa_\omega(\omega_1))=(\Lambda_\omega'(\omega_2),\kappa'_\omega(\omega_2))
\label{first_equality}
\end{equation}
and 
\begin{equation}
(\Lambda_\omega(\omega_2),\kappa_\omega(\omega_2))=(\Lambda_\omega'(\omega_1),\kappa'_\omega(\omega_1))
\label{second_equality}
\end{equation}
Here, we have used the definition $(\Lambda_\omega(\omega_i),\kappa_\omega(\omega_i)) := (v_\omega(\omega_i) \tau_\omega(\omega_i), C_\omega(\omega_i) v_\omega^2(\omega_i) \tau_\omega(\omega_i)/3)$, for $i=1, 2$ (and similarly for the $(\Lambda'_\omega(\omega_i),\kappa'_\omega(\omega_i))$ pair).

The above can be shown by writing
\begin{equation}
 \tau'_\omega(\omega_1) v_\omega(\omega_1)= \left [ \frac{ \tau_\omega(\omega_2) v_\omega(\omega_2)}{ v_\omega(\omega_1)} \right] v_\omega(\omega_1) = \tau_\omega(\omega_2) v_\omega(\omega_2) ,
\label{MFP_rel}
\end{equation}
where we have used equation \eqref{taus_rel} to obtain the second relationship. The heat conductivity value corresponding to this free path is
\begin{align}
\frac{1}{3} C_\omega(\omega_1) v_\omega^2(\omega_1) \tau'_\omega(\omega_1) &= \frac{1}{3} C_\omega(\omega_1) v_\omega(\omega_1) \left[ v_\omega(\omega_1) \tau'_\omega(\omega_1) \right]\nonumber \\ &= \frac{1}{3} C_\omega(\omega_2) v_\omega(\omega_2) \left[ v_\omega(\omega_2) \tau_\omega(\omega_2) \right] \nonumber \\ &=\frac{1}{3} C_\omega(\omega_2) v_\omega^2(\omega_2) \tau_\omega(\omega_2) ,
\label{k_rel}
\end{align}
where we have used both the assumption $v_\omega(\omega_2)\neq v_\omega(\omega_1)$ and equation \eqref{MFP_rel} to obtain the second equality. Relations \eqref{MFP_rel} and \eqref{k_rel} are equivalent to relation \eqref{second_equality}. 

Using the same process as \eqref{MFP_rel} and \eqref{k_rel} for $\omega_2$ (replacing $\omega_1$ by $\omega_2$), we obtain relation \eqref{first_equality}. 

These results (relations \eqref{first_equality} and \eqref{second_equality}) imply  that $\mathfrak{f}(\Lambda)=\mathfrak{f}'(\Lambda)$ for all free paths while there are frequencies for which $\tau_\omega(\omega) \neq \tau'_\omega(\omega)$ $\blacksquare$.

\section*{Appendix B: Silicon material properties}
\begin{figure}[H]
\centering
\begin{subfigure}{.48\textwidth}
\centering
\includegraphics[width=1.0\linewidth]{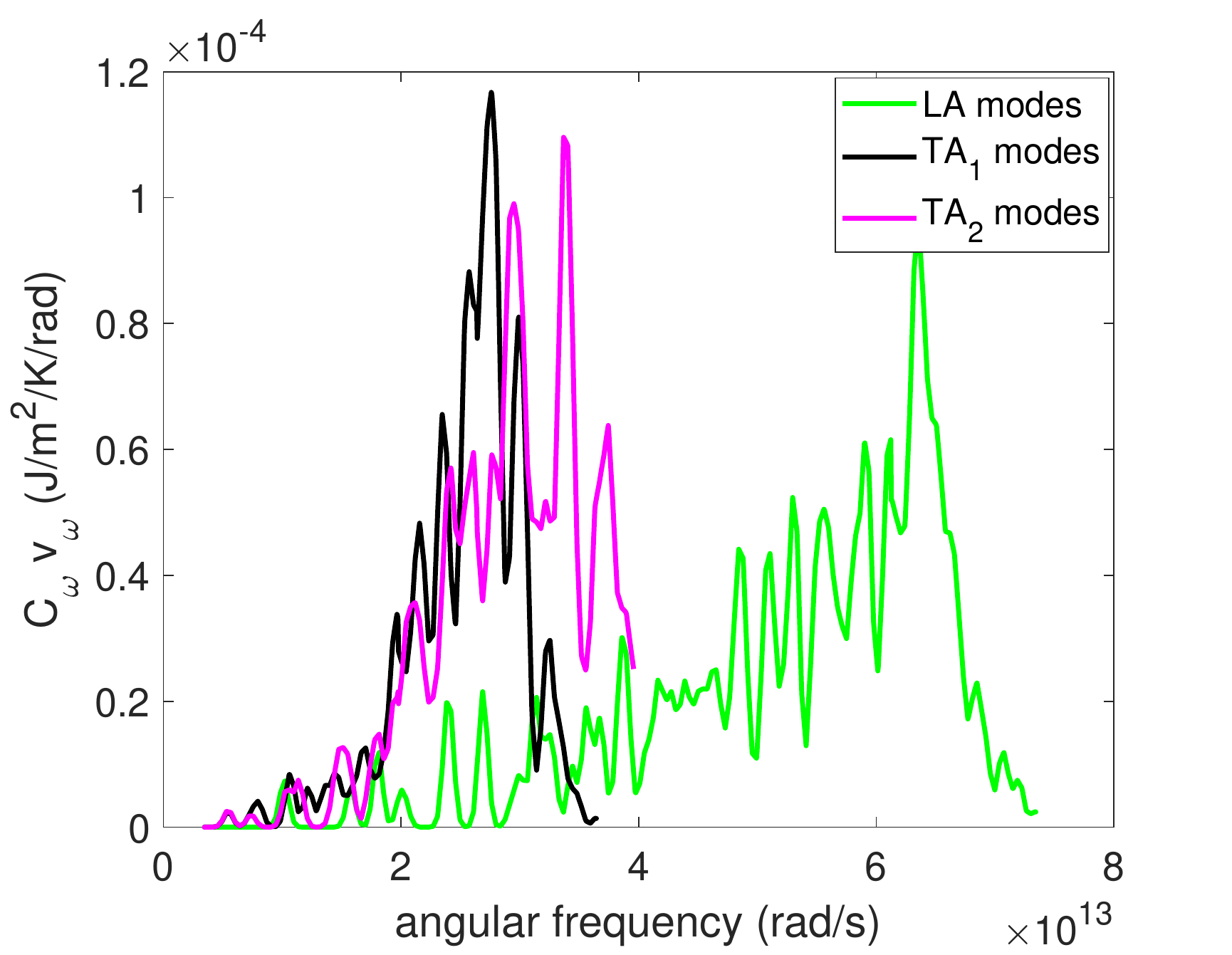}
\caption{}
\label{ab_initiomodell}
\end{subfigure}
\begin{subfigure}{.48\textwidth}
\centering
\includegraphics[width=1.0\linewidth]{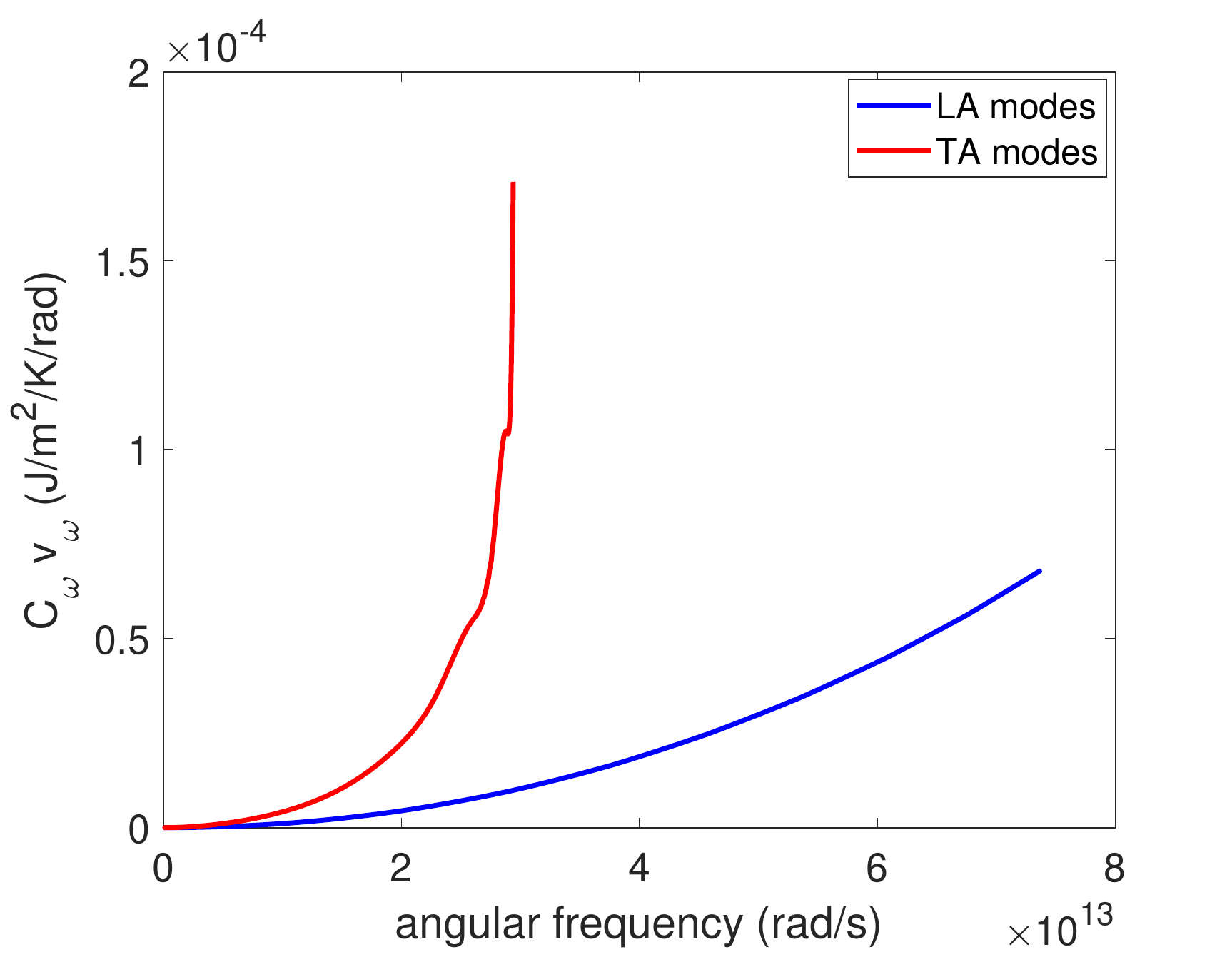}
\caption{}
\label{Hollandmodell}
\end{subfigure}
\caption{The plot of $C_\omega v_\omega$ versus frequency for the {\it ab initio} (a) and the Holland model (b).}
\label{MFP_unique}
\end{figure}

Figure \ref{MFP_unique} shows the plot of $C_\omega v_\omega$ for two different sets of DFT-based silicon material properties, referred to as the Holland and the {\it ab initio} models \cite{Mojtaba2016, Mojtaba2018, MojtabaAPL, MojtabaThesis}. Longitudinal acoustic and transverse acoustic branches are denoted by $LA$ and $TA$, respectively. We observe that in both cases, considering all branches, a number of modes have the same $C_\omega v_\omega$ value for different values of $\omega$. In particular, for the {\it ab initio} model, even considering one branch, there are many modes that have the same $C_\omega v_\omega$ value. The existence of many such modes implies that a number of other relaxation-time distributions exist which could have the same $F(\Lambda)$ as the {\it ab initio} model. Note that the results provided in section~\ref{numer_dem} are based on Holland model.

\section*{Appendix C: Bayesian inference}
The likelihood function that we considered in our study is defined as 
\begin{equation}
p\Big(\{ T_\text{m}(t, {\bf x}=0; L)\}_{t,L} \big| {\bf U}\Big)= {\cal N} \left( \{ T_\text{m}(t, {\bf x}=0; L)\}_{t,L}; \{ T_\text{BTE}(t, {\bf x}=0; L)\}_{t,L}, \sigma_m^2 \right) ,
\label{noise_model}
\end{equation}
in which $\sigma_\text{m}=0.01$ K. The prior distribution $\pi({\bf U})$ of the parameter {\bf U} --- its distribution before the observations--- here is taken to be a Laplace distribution,
\begin{equation}
\pi({\bf U})= \prod_{i} \frac{1}{2{\bf \Sigma}_i}\exp{\left( -\frac{|{\bf U}_i- {\bf \mu}_i|}{{\bf \Sigma}_i} \right)} ,
\end{equation}
whose variance for each component ${\bf U}_i$ is $2{\bf \Sigma}_i^2$ and $\mu_i$ is its mean (corresponding to the correct solution). The variance and mean values for different components are taken to be similar to those in \cite{MojtabaThesis}. Here, $\{ T_\text{m}(t, {\bf x}=0; L)\}_{t,L}$ are the set of all temperature measurements (observations). Equation \eqref{noise_model} implies that here we are assuming that the temperature measurements are corrupted by a noise of Gaussian form with the standard deviation of $\sigma_\text{m}=0.01$ K. The value of $\sigma_m$ is consistent with the noise in the common thermal spectroscopy experiments \cite{MojtabaAPL}. The choice of Laplace distribution for the prior is motivated by two observations:

\begin{itemize}
\item Here we are more interested in the sensitivity of the reconstructed property, the relaxation-times function, to the noise in the temperature measurement. Therefore, instead of the least-informative uniform distribution, we are assuming a prior with higher probability around the correct solution. This is motivated by our observation that this optimization (global optimization as well as optimizations in \cite{Mojtaba2016, Mojtaba2018, MojtabaAPL}) converges to a solution close to the correct function. In other words, the Bayesian analysis is a complement to the uniqueness analyses, in the sense that while the results provided in figure \ref{GNM_plot} show that the objective function has a global minimum, in the Bayesian analysis we study how sensitive the objective function is to the measurement noise in the vicinity of that global minimum.
\item Due to our choice of objective function ${\cal L}$ ($L$-1 norm) we used here and previously in \cite{Mojtaba2016, Mojtaba2018, MojtabaAPL}, the Laplace distribution is expected to be a more accurate representation of the behavior of the objective function compared to other common distributions such as Gaussian distribution (compare the similarity between Laplace distribution and the shape of the objective function at the minimum in the supplementary document of \cite{MojtabaAPL}). The result corresponding to a Gaussian distribution counterpart of this distribution can be found in \cite{MojtabaThesis}  
\end{itemize}

The posterior distribution was sampled using the Metropolis-Hasting (MH) algorithm \cite{Metro}. The parameters of the algorithm (e.g., the variance of the proposal distribution) are similar to those of \cite{MojtabaThesis}. The acceptance rate for the chosen proposal distribution is 0.21 which is close to the optimal value of 0.23 for multi-dimensional variables \cite{accept_rate}. We have used a total of about $N_\text{MH}= 7\times10^5$ sample points; the first $10\%$ of all samples, approximately, were used as burn-in samples.

\end{document}